\documentclass[final,1p,times,twocolumn,authoryear]{elsarticle}

\usepackage{amssymb}

\usepackage{lineno}

\usepackage{graphicx}
\usepackage{txfonts}
\usepackage{natbib}   
\usepackage{amstext}
\usepackage{mathabx}
\usepackage{multirow}
\usepackage{ulem}
\usepackage{color}

\newcommand{\jgr}{J. Geophys. Res. }
\newcommand{\grl}{Geophys. Res. Lett. }
\newcommand{\icarus}{Icarus }
\newcommand{\aap}{Astron. Astrophys. }
\newcommand{\apj}{Astrophys. J. }
\newcommand{\apjl}{Astrophys. J. Lett. }
\newcommand{\apjs}{Astrophys. J. Suppl. }
\newcommand{\aj}{Astron. J. }
\newcommand{\planss}{Planet. Space Sci. }

\newcommand{\ssr}{Space Sci. Rev. }
\newcommand{\mnras}{Mon. Not. R. Astron. Soc. }
\newcommand{\nat}{Nature }
\newcommand{\prb}{Phys. Rev. B }
\newcommand{\expastr}{Exp Astron. }
\newcommand{\gca}{Geochim. Cosmochim. Acta }
\newcommand{\jchemphys}{J. Chem. Phys. }
\newcommand{\jqsrt}{J. Quant. Spectrosc. Radiat. Transfer }
\newcommand{\pasp}{Publ. Astron. Soc. Pac }

\newcommand{\yco}{$y_{\mathrm{CO}}$}
\newcommand{\y}[1]{$y_{\mathrm{#1}}$}

\newcommand{\Flux}[1]{$\phi_{\mathrm{#1}}$}

\newcommand{\Fluxunit}{\,cm$^{-2}\cdot$s$^{-1}$}
\newcommand{\Kunit}{\,cm$^{2}\cdot$s$^{-1}$}

\newcommand{\dix}[1]{$\times10^{#1}$}
\newcommand{\fig}[1]{Fig.~\ref{#1}}
\newcommand{\figs}[2]{Figs.~\ref{#1} and~\ref{#2}}

\newcommand{\tab}[1]{Table~\ref{#1}}

\newcommand{\derive}[2]{\frac{d{#1}}{d{#2}}}

\journal{Icarus}

\begin{document}

\begin{frontmatter}



\title{Thermochemistry and vertical mixing in the tropospheres of Uranus and Neptune: How convection inhibition can affect the derivation of deep oxygen abundances}


\author[inst1]{T.~Cavali\'e\corref{cor1}}
\author[inst2,inst3]{O.~Venot}
\author[inst4]{F.~Selsis}
\author[inst4]{F.~Hersant}
\author[inst5]{P.~Hartogh}
\author[inst4]{J.~Leconte}

\cortext[cor1]{Corresponding author. E-mail address: thibault.cavalie@obspm.fr (T. Cavali\'e)}

\address[inst1]{LESIA, Observatoire de Paris, PSL Research University, CNRS, Sorbonne Universités, UPMC Univ. Paris 06, Univ. Paris Diderot, Sorbonne Paris Cité, F-92195 Meudon, France}
\address[inst2]{Instituut voor Sterrenkunde, Katholieke Universiteit Leuven, Leuven, Belgium}
\address[inst3]{Laboratoire Interuniversitaire des Syst\`emes Atmosph\'eriques (LISA), CNRS UMR 7583, Universit\'e Paris-Est Cr\'eteil, Universit\'e Paris Diderot, 61 avenue du G\'en\'eral de Gaulle, 94010 Cr\'eteil Cedex, France}
\address[inst4]{Laboratoire d'Astrophysique de Bordeaux, Univ. Bordeaux, CNRS, B18N, allée Geoffroy Saint-Hilaire, 33615 Pessac, France}
\address[inst5]{Max-Planck-Institut f\"ur Sonnensystemforschung, 37077 G\"ottingen, Germany}

\begin{abstract}
Thermochemical models have been used in the past to constrain the deep oxygen abundance in the gas and ice giant planets from tropospheric CO spectroscopic measurements. Knowing the oxygen abundance of these planets is a key to better understand their formation. These models have widely used dry and/or moist adiabats to extrapolate temperatures from the measured values in the upper troposphere down to the level where the thermochemical equilibrium between H$_2$O and CO is established. The mean molecular mass gradient produced by the condensation of H$_2$O stabilizes the  atmosphere against convection and results in a vertical thermal profile and H$_2$O distribution that departs significantly from previous estimates. We revisit O/H estimates using an atmospheric structure that accounts for the inhibition of the convection by condensation. We use a thermochemical network and the latest observations of CO in Uranus and Neptune to calculate the internal oxygen enrichment required to satisfy both these new estimates of the thermal profile and the observations. We also present the current limitations of such modeling.  
\end{abstract}


\end{frontmatter}


Received 9 December 2016 ; Revised 3 March 2017 ; Accepted 10 March 2017

DOI: 10.1016/j.icarus.2017.03.015

\section{Introduction}
One of the great mysteries in the Solar System is how the gas and ice giant planets formed from the protoplanetary disk 4.5 billion years ago. This question is even more relevant regarding ice giants after the discovery of the commonality of Neptune-class planets among the exoplanets detected by Kepler \citep{Batalha2013,Fressin2013}. Two scenarios have been proposed regarding the formation of giant planets: disk gravitational instability \citep{Boss1997} and core accretion \citep{Pollack1996}. These scenarios differ not only in the time required to form planets (a few hundred years vs. several million years, respectively), but also in the final composition of the planets' interiors. While gravitational instability should result in $\sim$solar abundances of heavy elements (except if a significant external source of heavy elements is incorporated after the planet formation), core accretion formation should lead to an enrichment in heavy elements increasing with heliocentric distance. The level of enrichment then would depend on how the ices of the planetesimal that formed the cores of these planets condensed. Here again, several competing scenarios exist: condensation in amorphous ices \citep{Bar-Nun1988,Owen1999} and clathration \citep{Lunine1985,Gautier2001,Hersant2004,Gautier2005}. 

While measuring the D/H ratio in the giant planets \citep{Lellouch2001,Feuchtgruber2013} gives us an insight on the origin and condensation temperature of the protoplanetary ices in the outer Solar System \citep{Hersant2001}, measuring the heavy element abundances can help constraining the ice condensation processes. Enrichment in heavy elements has been observed by the Galileo probe in Jupiter's troposphere, with an average enrichment factor of $\sim$4$\pm$2 in C, N, S, Ar, Kr and Xe, except for O which was found depleted  \citep{Niemann1998,Atreya1999,Mahaffy2000,Wong2004,Owen2006}. The O measurement may only be a lower limit because the Galileo probe most likely entered a dry zone of Jupiter and did not reach the levels where H$_2$O is well-mixed \citep{Wong2004}, though alternative scenarios explaining an oxygen depletion exist \citep{Lodders2004,Mousis2012}. The global enrichment in heavy elements reported by Galileo favors the core accretion scenario for Jupiter. This formation model, if applied to the other giant planets, predicts enrichment factors that increase with increasing heliocentric distance: $\sim$7, 30 and 50 for Saturn, Uranus, and Neptune, respectively \citep{Owen2003}. At Saturn, the C, N, P and S abundances have been measured \citep{Fletcher2009,Hersant2008,Mandt2015} and are also found enriched compared to the solar value. At Uranus and Neptune, the information is more sparse, with only the C abundance has constraints \citep{Baines1995,Sromovsky2008,Karkoschka2009}. 

The key measurement that would enable differentiating the condensation processes of the planetesimal ices and hence how other heavy elements were trapped is the deep water abundance. Indeed, the clathration scenario needs a larger amount of water at the time of the condensation of the planetesimal ices than the amorphous ice scenario \citep{Owen2007}, especially if the efficiency of the clathration process was lower than 100\%. While Galileo probably failed to measure the Jovian deep water abundance below the water cloud \citep{Niemann1998} because it entered a dry hot spot and probably did not reach the base of the water cloud, Juno \citep{Matousek2007} should shed light on this long lasting question. However, there is no such mission planned in the near future to measure the deep water abundance in the other giant planets. A few mission concepts are being developed for Saturn \citep{Mousis2012,Mousis2014,Atkinson2016} and the ice giants \citep{Arridge2012,Arridge2014,Masters2014}, but these challenging missions require probes that would have to survive high pressures to reach below the water cloud (up to $\sim$100 bars in the ice giants). In principle, cm-waves can probe down to several tens of bars and could probe below the water cloud \citep{dePater1991,Hofstadter2003,Klein2006}. However, the opacity at these levels can also be caused by NH$_3$ and H$_2$S and the degeneracy is difficult to waive (e.g., \citealt{dePater2005}). Therefore, it is important to find other ways to constrain the deep water abundance. One interesting way is to take advantage of the chemical quenching of species like CO. The observed abundance of CO is in chemical disequilibrium and is inherited from deeper layer, where its abundance is in thermochemical equilibrium with water, through the combination of fast vertical mixing and slow chemical kinetics. 

Following the first detection of CO in the atmosphere of Jupiter by \citet{Beer1975}, a simple model was proposed to constrain the deep water abundance in Jupiter by studying the tropospheric thermochemistry and vertical transport, and in particular the following thermochemical equilibrium reaction:
  \begin{equation}
    \mathrm{H}_2\mathrm{O}+\mathrm{CH}_4=\mathrm{CO}+3\mathrm{H}_2. \label{equilibrium_reaction}
  \end{equation}
This model, first developed by \citet{Fegley1988} is based on the approximation that the tropospheric mole fraction of CO is fixed at a so-called ``quench'' level, where the chemical timescale of the conversion of CO into H$_2$O becomes longer than the timescale for its vertical transport by convection. This kind of model relies on the determination of the rate-limiting reaction of the conversion scheme. Therefore, assuming the kinetics of this rate-limiting reaction is known, the kinetics of the whole conversion scheme is constrained and the measured upper tropospheric mole fraction of CO can be linked to the deep water abundance. \citet{Prinn1977} initially proposed this reaction to be H$_2$CO$+$H$_2$$\rightarrow$CH$_3$$+$OH. Later, \citet{Yung1988} proposed a two-step reaction scheme in which the rate-limiting reaction was H$+$H$_2$CO$+$M$\rightarrow$CH$_3$O$+$M. This ``quench-level'' approximation model was then used by \citet{Lodders1994} to constrain the atmospheric O/H ratio in all giant planets. However, \citet{Smith1998} showed that the assumptions of these modelers on diffusion timescales were incorrect. His work was then applied by \citet{Bezard2002} to Jupiter, by \citet{Visscher2005} and \citet{Cavalie2009} to Saturn, and by \citet{Luszcz-Cook2013} to Neptune using either the \citet{Prinn1977} or \citet{Yung1988} limiting reactions, which renders any comparison between these results hazardous. 

\citet{Visscher2010} first applied a comprehensive thermochemical and diffusive transport model to the troposphere of Jupiter. They have evaluated the Jovian deep water mole fraction to be (0.25-6.0)\dix{-3}, corresponding to an enrichment of 0.3-7.3 times the protosolar abundance (9.61\dix{-4}). Similar thermochemical models have been applied ever since to investigate the thermochemistry in exoplanets \citep{Visscher2011,Moses2011,Moses2013,Venot2012,Venot2014,Venot2015} and in Solar System giants \citep{Cavalie2014,Mousis2014,Wang2016}. While all these models improve on the modeling of deep tropospheric chemistry compared to the ``quench-level'' approximation studies, they still have to rely on assumptions made on tropospheric vertical mixing and temperatures. Interestingly, recent theoretical work enables progress in the determination of these quantities: 
\begin{itemize}
   \item Vertical mixing: vertical mixing in the troposphere of giant planets is caused by convection, and it is usually modeled by an eddy diffusion coefficient. It is estimated within an order of magnitude from the mixing length theory \citep{Stone1976}. Following rotating tank experiments, a recent paper by \citet{Wang2015} proposes a new formulation to estimate this coefficient with a narrower error bar. The authors even show that this vertical mixing is latitude-dependent, with a stronger magnitude at low latitudes. 
   \item Tropospheric temperatures extrapolation: Until now, dry and/or moist adiabats have been used to extrapolate the thermal profiles of Solar System giant planets from observations around 1-bar to deeper levels (e.g., \citealt{Luszcz-Cook2013}). In a new paper, \citet{Leconte2017} propose a new criterion to compute the thermal gradient in the giant planet tropospheres. It takes into account not only dry and moist processes, but also the effect of the mean molecular weight gradient associated with the condensation of H$_2$O and can produce a jump in temperature at the H$_2$O condensation level that is caused by a thin radiative layer.
\end{itemize}
Both are composition-dependent and remain therefore quite uncertain. In addition, the possible radiative gradient at the H$_2$O condensation level is poorly constrained and the CH$_4$ tropospheric abundance is only known within a factor of two in Uranus and Neptune \citep{Baines1995,Sromovsky2008,Karkoschka2009}. We have therefore chosen to study the parameter space -- which consists of the following four parameters: O/H, C/H, temperature, $K_{zz}$ -- with a thermochemical and diffusion model of Uranus and Neptune to better evaluate the parameter space that is compliant with observations of CO in their upper tropospheres.

In this paper, we apply to the Solar System Ice Giants the thermochemical and diffusion model of \citet{Venot2012} along with the prescriptions of \citet{Leconte2017} to compute tropospheric thermal profiles. We present our models in Section 2, and review the various observational constraints on composition and temperature in Section 3. We detail the results of the 4D parameter space simulations and our nominal case in Section 4. We discuss our results in Section 5, and detail other sources of uncertainties that currently limit these kind of models in Section 6. We give our conclusions in Section 7.

\section{Models \label{Models}}
We have adapted the thermochemical and diffusion model of \citet{Venot2012}, initially developed for the atmospheres of warm exoplanet atmospheres, to the giant planet tropospheres to constrain their deep oxygen abundance from CO observations. Obtaining the tropospheric composition from thermochemical and diffusion calculations requires some knowledge on the troposphere temperature, composition, and vertical mixing. In the following sections, we illustrate how we have run sequentially various steps to eventually link the observed tropospheric CO abundances to the deep oxygen abundances in Uranus and Neptune. These steps include:
\begin{enumerate}
  \item temperature profile extrapolation to deep levels where thermochemistry prevails, assuming a given composition of the main compounds (H$_2$, He, CH$_4$ and H$_2$O)
  \item thermochemical equilibrium calculation using the temperature profile and a given elemental composition, to derive an initial composition state for the next round of computations 
  \item thermochemistry and diffusion calculations using the temperature profile and initial composition state of step 1 and step 2 and assuming a given vertical mixing $K_{zz}$, to obtain the steady state tropospheric composition
  \item $K_{zz}$ assumption cross-check with theoretical estimates
\end{enumerate}

To run the thermochemistry and diffusion model, we need at minimum to set as initial conditions the internal elemental abundances of H, He, C, N, and O, a thermal profile, and an eddy diffusion coefficient. The case of N will not be discussed any further as nitrogen chemistry has no significant impact on carbon and oxygen chemistries \citep{Cavalie2014}, the deep oxygen abundance is an assumption of the model, and other elemental abundances can be estimated a priori. 

  \subsection{Step 0 - Estimating the internal composition by neglecting chemistry}
  As will be shown in what follows, the computation of the thermal profile needs a priori knowledge of the deep composition in terms of H$_2$, He, CH$_4$ and H$_2$O. To estimate these molecular abundances as well as the elemental abundances of H, He and C, we start from the upper tropospheric abundances of H$_2$, He, CH$_4$, and an initial condition on the deep O abundance. Assuming that chemistry and the abundance of compounds other than H$_2$, He, CH$_4$, and H$_2$O, can be neglected, and that the upper tropospheric abundance of H$_2$O is negligible (because of condensation), we can derive a set of equations to link the upper tropospheric and deep tropospheric composition:
\begin{equation}
    X_{\mathrm{He}} = y_{He}^{\mathrm{top}} / (y_{\mathrm{H}_2}^{\mathrm{top}} + 2 y_{\mathrm{CH}_4}^{\mathrm{top}}) \times (X_{\mathrm{H}} - 2 X_{\mathrm{O}}) / 2, \label{eq_X_He}
\end{equation}
\begin{equation}
    X_{\mathrm{C}} = X_{\mathrm{He}} \times y_{\mathrm{CH}_4}^{\mathrm{top}}/y_{\mathrm{He}}^{\mathrm{top}}, \label{eq_X_C}
\end{equation}
\begin{equation}
    y_{\mathrm{H}_2\mathrm{O}}^{\mathrm{bottom}} = X_{\mathrm{O}} / (X_{\mathrm{H}}/2 + X_{\mathrm{He}} - X_{\mathrm{C}}), \label{eq_y_H2O}
\end{equation}
where $X_{i}$ is the abundance of element $i$, $y_{i}^{\mathrm{top}}$ is the mole fraction of compound $i$ in the upper of the troposphere (but below the CH$_4$ cloud), and $y_{i}^{\mathrm{bottom}}$ is the mole fraction of compound $i$ in the deep troposphere. By convention, $X_\mathrm{H}=10^{12}$.
  
  In practice, we have measurements from which we can determine $y_{\mathrm{He}}^{\mathrm{top}}$, $y_{\mathrm{CH}_4}^{\mathrm{top}}$, and $y_{\mathrm{H}_2}^{\mathrm{top}}$ (see Section~\ref{Observation_constraints}), and we assume a deep oxygen abundance $X_\mathrm{O}$. We can thus derive an estimate of $X_\mathrm{He}$, $X_\mathrm{C}$, and $y_{\mathrm{H}_2\mathrm{O}}^{\mathrm{bottom}}$, which are needed for the preparation of the thermochemistry and diffusion computations, by using sequentially equations \ref{eq_X_He}, \ref{eq_X_C}, and \ref{eq_y_H2O}. We control the validity of these assumptions on $X_\mathrm{He}$ and $X_\mathrm{C}$ a posteriori, by checking that the final upper tropospheric abundances of He and CH$_4$ fit the observations.

  \subsection{Step 1 - Extrapolating tropospheric temperatures}
  Formation models of the giant planets of the Solar System predict enrichment factors for heavy elements that increase together with heliocentric distance. The precise values for the O enrichment factor depend on the ice condensation scenario (amorphous ices vs. clathration). But even in the case where the O enrichment should be the lowest, i.e., in the scenario where ices condense and trap volatiles  in amorphous form, the expected enrichment factors are 4, 7, 30, and 50 (respectively) for Jupiter, Saturn, Uranus, and Neptune (respectively), according to \citet{Owen2003}. The high abundance of tropospheric CO in Neptune could even indicate that O is enriched by a factor $>$100 \citep{Lodders1994,Luszcz-Cook2013}. Actually, above an enrichment factor of $\sim$100-150 in Uranus and Neptune, we expect to see the abundance of O exceeding the abundance of He. Given the molecular mass of the main O carrier (i.e., water) compared to the molecular mass of the main atmospheric constituents (i.e., H$_2$ and He), a significant mean molecular mass gradient $\nabla_\mu$ is expected around the layers where water condenses. Consequently, the Ledoux stability criterion \citep{Ledoux1947,Sakashita1959} must be used instead of the Schwarzschild stability criterion in the computations of the temperature gradient $\nabla_\textrm{T}$, as double diffusive processes may appear. The temperature gradient can be significantly affected, as shown by \citet{Guillot1995} in the case of the condensation of methane. 
  
  Hereafter, we will especially consider the cases of Uranus and Neptune, as these planets are the most ``symptomatic'' cases of mean molecular weight discontinuities around the water condensation level, because of the expected large oxygen enrichments. 
    
    \subsubsection{Methodology}
    The thermochemistry and diffusion computations require a thermal profile for the troposphere down to the region where thermochemistry prevails over vertical diffusion, i.e. below the CO quench-level in our case. 
    
    Previous studies of the Solar System giant planets thermochemistry have used a variety of dry and/or moist adiabats, based on the application of the \citet{Schwarzschild1958} stability criterion, to compute the tropospheric thermal structures (e.g., \citealt{Lodders1994,Luszcz-Cook2013}). In a new paper, \citet{Leconte2017} show that, in hydrogen-rich  atmospheres, the mean molecular weight gradient around the cloud base (of a species heavier than H$_2$) can be strong enough to stabilize the atmosphere against convection  (i.e. inhibit moist convection) in this region. Although it shares some similarities with the processes described by Ledoux that work for non condensable species \citep{Ledoux1947,Sakashita1959}, this mechanism can suppress moist convection when the enrichment in the condensable species is higher than a critical threshold. Because the interior still needs to release its energy, a stable radiative layer with a steep temperature gradient develops. \citet{Guillot1995} has already shown that such an effect can be produced by CH$_4$ in Uranus and Neptune. We can expect this effect to happen with H$_2$O too in Uranus and Neptune, in the transition zone between the H$_2$O-rich region (deep troposphere) and the H$_2$O-poor region (upper troposphere), and we have therefore implemented their stability criterion when extrapolating thermal profiles. We will refer to ``3-layer profiles'' for thermal profiles extrapolated using the prescription of \citet{Leconte2017}.
    
    In the computations we present in this paper, we start from the temperature measured at the 2-bar level (see Table~\ref{Obs_list}). We will explore various cases for extrapolating deep tropospheric thermal profiles: (i) dry processes leading to a dry adiabat, (ii) latent heat effects leading to a moist adiabat, (iii) latent heat and mean molecular weight effects leading to what we will refer to as ``3-layer profiles''. Finally, because magnitude of the temperature jump in the radiative layer of the 3-layer profiles is quite uncertain, we derive an extreme case (iv) from the \citet{Leconte2017} formulations in which the temperature jump of the radiative layer is limited by the deep water abundance.

    \subsubsection{Dry adiabat}
    Starting from the 2-bar level temperature, we extrapolate to deeper levels using the \citet{Schwarzschild1958} criterion, which translates into the following:
    \begin{equation}
      T_{i+1}=T_i\times \exp\left(\nabla_{\mathrm{dry}}\frac{\ln(p_{i+1})}{\ln(p_{i})}\right)
      \label{dry}
    \end{equation}
    and 
    \begin{equation}
      \nabla_{\mathrm{dry}} = \frac{R}{\mu c_\mathrm{p}},
      \label{nabla_dry}
    \end{equation}
    where $\mu$ is the mean molecular weight, and $R$ is the ideal gas constant. We use the temperature-dependent expressions given by the NIST for the specific heat capacities of He, CH$_4$ and H$_2$O. For H$_2$, we use the temperature- and pressure-dependent data from the Cryogenic Data Handbook\footnote{https://www.bnl.gov/magnets/staff/gupta/cryogenic-data-handbook/Section3.pdf}. We assume equilibrium hydrogen, in agreement with observations of \citet{Baines1995} of the upper tropospheres of Uranus and Neptune. Because we start from the 2-bar level in both Uranus and Neptune, the lowest temperatures we consider are sufficiently high so that we do not have to consider the effects of CH$_4$ condensation \citep{Guillot1995}. We then compute the next levels with very small steps in $\ln(p)$ ($\sim10^{-4}$), with $p$ in Pa.
     
    This kind of profile is obviously the least realistic given the expected high abundance of H$_2$O in the interiors of Uranus and Neptune.

    \subsubsection{Moist adiabat}
    In the region where H$_2$O condenses, latent heat is released and the gas is heated. Convection is thus reinforced. Adding this effect to the previous stability criterion results in replacing $\nabla_{\mathrm{dry}}$ in equation~(\ref{dry}) by the expression given in \citet{Leconte2013}:
    \begin{equation}
      \nabla_{\mathrm{moist}} = \frac{p}{p-p_\textrm{v}} \frac{(1-q_\textrm{v})R_\textrm{a} +  \frac{q_\textrm{v}L_\textrm{v}}{T}}
{q_\textrm{v} c_\textrm{p,v} + q_\textrm{a} c_\textrm{p,a} + q_\textrm{c} c_\textrm{p,c} 
+ q_\textrm{v} \frac{L_\textrm{v}}{T} \frac{p}{p-p_\textrm{v}} \frac{\mathrm{d}\ln p_\textrm{s}}{\mathrm{d}\ln T}},
      \label{nabla_moist}
    \end{equation}
    where subscripts $i$$=$a,v,c refer to non-condensable gas, condensable gas, and condensed material (solid or liquid), respectively. $p_i$ is the partial pressure, $M_i$ the molar mass, $q_i$ is the mass mixing ratio and is the ratio of the mass of component $i$ over the mass of the gas, $c_{\textrm{p},i}$ the mass heat capacity, and $R_i=R/M_i$. $p_s$ is the saturation vapor pressure (which is equal to $p_\textrm{v}$ in our case). For H$_2$O, we use the Tetens formula. $L_\mathrm{v}$ is the latent heat of vaporization. We assume here that all the condensed material falls into deeper layers. As a consequence, the term $q_\textrm{c} c_\textrm{p,c}$ disappears and the process can no longer be considered as strictly adiabatic (hence the term \textit{pseudo-adiabat} sometimes used). We note that below the H$_2$O cloud, $\nabla_{\mathrm{moist}}=\nabla_{\mathrm{dry}}$.

    \subsubsection{3-layer profiles}
    Condensation of H$_2$O can result in a significant gradient in molecular weight when high oxygen enrichments are considered. The increase of molecular weight with pressure introduces a stabilizing effect. \citet{Leconte2017} derived a new stability criterion that applies in this situation. When
    \begin{equation}
      \varpi q_\mathrm{v} \frac{\mathrm{d}\ln p_\mathrm{s}}{\mathrm{d} \ln T} > 1,
      \label{Ledoux_criterion}
    \end{equation}
    where $\varpi\equiv(\mu_v-\mu_a)/\mu_v$ and $q_\mathrm{v}\equiv q_{\mathrm{H}_{2}\mathrm{O}}$, convection is inhibited and a radiative layer is formed. This layer is stable against double-diffusive processes in 1D, as demonstrated by \citet{Leconte2017}. The thermal gradient from (\ref{nabla_moist}) must then be replaced by a radiative gradient
    \begin{equation}
      \nabla_\textrm{r} = \frac{3}{16} \frac{\kappa p T_\textrm{int}^4}{g T^4},
      \label{nabla_rad}
    \end{equation}
    where $\kappa$ is the Rosseland mean opacity, $g$ the gravity and $T_\textrm{int}$ the temperature associated with the internal heat flux (see Table~\ref{Obs_list}). For the Rosseland mean opacities, we use the parametrization from \citet{Valencia2013}. This parametrization requires a metallicity, which we compute by accounting for the elemental abundances of oxygen and carbon. 
    
\begin{figure}[htp]
  \begin{center}
  \includegraphics[width=10cm,keepaspectratio]{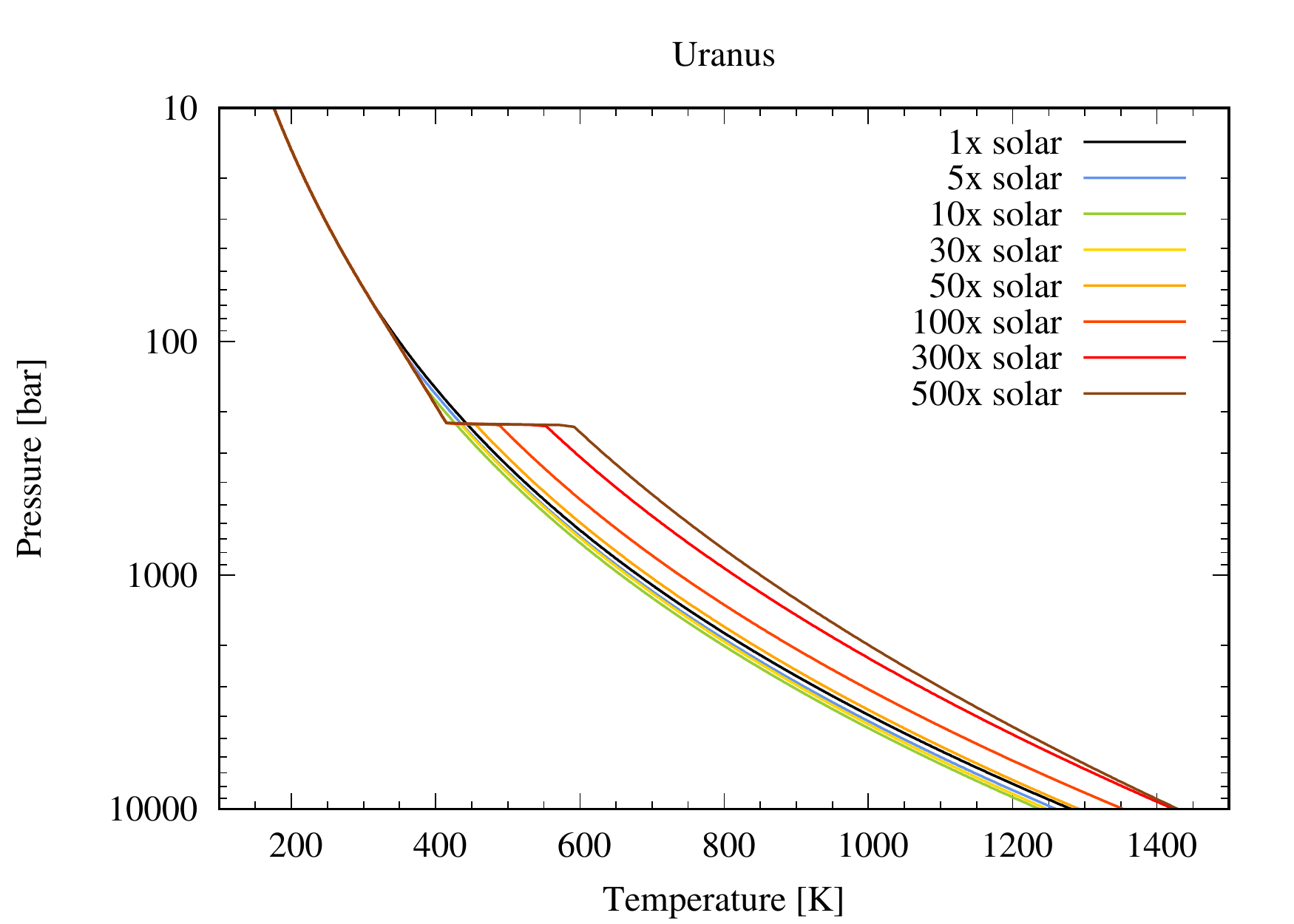}
  \caption{``3-layer'' thermal profiles computed for Uranus, assuming an O enrichment factor over the solar value ranging from 1 to 500. The case for which there is no water (not shown) would correspond to a dry adiabat, while low enrichment factors result in a wet adiabat. When the O abundance exceeds a threshold (O/H$\sim$20 times the solar value ; valid for both Uranus and Neptune), a significant jump in mean molecular weight translates into a significant jump in temperature in the layers where water condenses. In this thin layer ($\Delta z\sim$1\,km, whatever the O/H beyond the threshold), the convective temperature gradient is too high and the atmosphere is radiative. Refer to the online version for the color plot.}
  \label{Ledoux_Ura}
  \end{center}
\end{figure}

    Thus, if $q_{\mathrm{H}_{2}\mathrm{O}}$ exceeds a given threshold (see equation 17 in \citealt{Leconte2017}), a positive feedback appears in the condensation region. As the mean molecular weight gradient, that manifests itself as an increase of $q_{\mathrm{H}_{2}\mathrm{O}}$ in equation~(\ref{Ledoux_criterion}), stabilizes the layer, the thermal gradient can increase and exceed the moist gradient owing to (\ref{nabla_rad}). As a consequence, more H$_2$O is vaporized and the H$_2$O mole fraction at the level $i+1$ increases, resulting in an increase of the mean molecular weight gradient. For high enough oxygen elemental abundances, this creates a very localized radiative region ($\Delta z\sim$1\,km, whatever the O/H beyond the threshold) that separates the water-poor and water-rich regions. This interface is associated with a jump in temperature due to inefficient heat transport, as demonstrated by \citet{Leconte2017}. Profiles for various oxygen enrichments are displayed in \figs{Ledoux_Ura}{Ledoux_Nep} for Uranus and Neptune.

\begin{figure}[htp]
  \begin{center}
  \includegraphics[width=10cm,keepaspectratio]{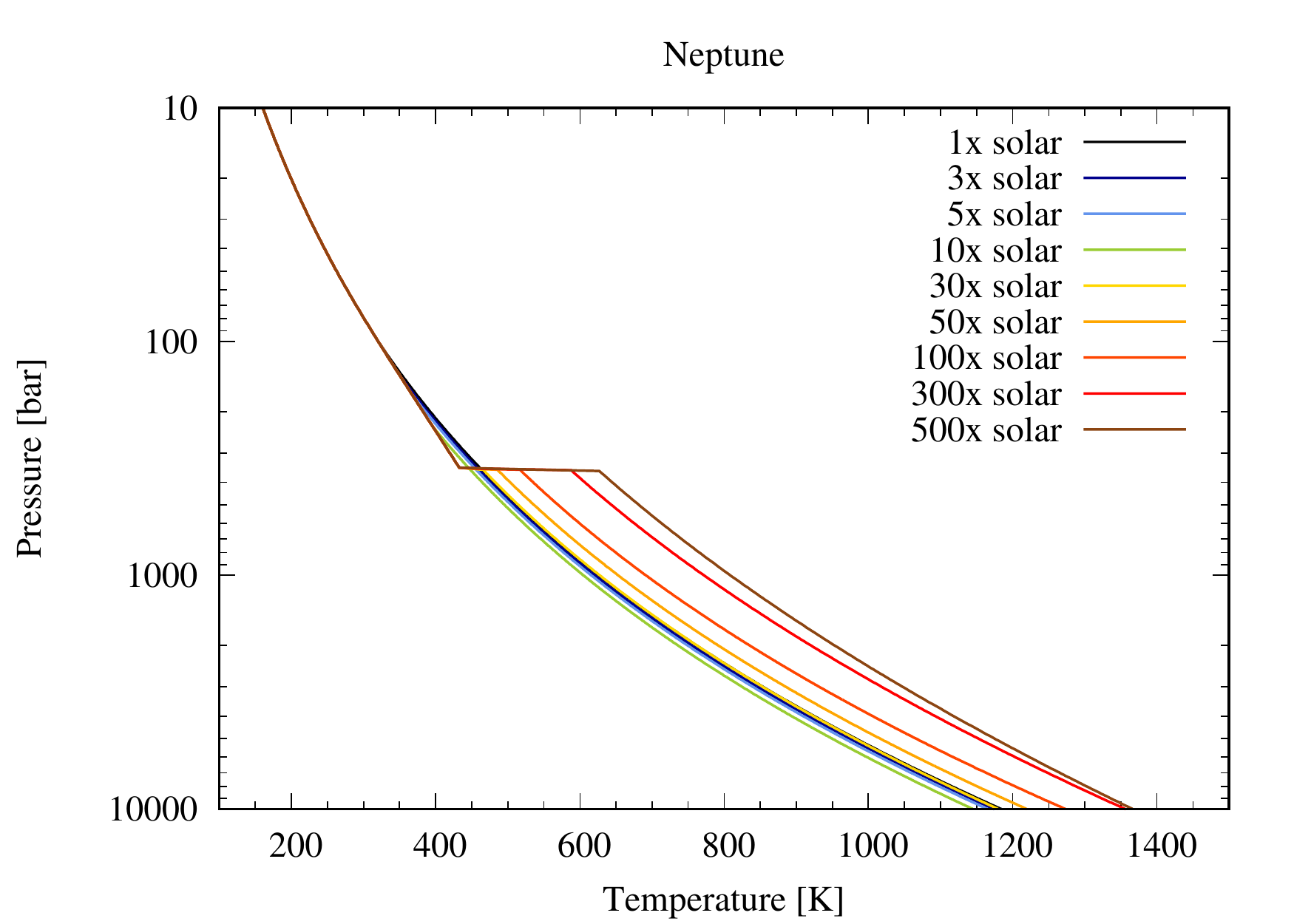}
  \caption{Same as \fig{Ledoux_Ura} for Neptune}
  \label{Ledoux_Nep}
  \end{center}
\end{figure}

    \subsubsection{Extreme case and sampling range}
    The magnitude of the temperature jump in the 3-layer profiles is controlled by the magnitude of the Rosseland opacities. Those are not well-constrained for heterogeneously enriched mixtures \citep{Valencia2013}. Moreover, any possible opacity of the condensate itself has been neglected with the formulation we have used. Because the jump in temperature at the transition zone between a water-rich and a water-poor region may be governed by these opacities, we may overestimate or underestimate this jump. 
    
    For illustration, let us take the example of Neptune around the nominal case that will be presented later in Section~\ref{Results}. Even if we divide the Rosseland opacities by a factor of 10, the upper tropospheric mole fraction of CO is unnoticeably altered. On the other hand, if the opacities are increased by a factor of 5, $y_{\mathrm{CO}}^{\mathrm{top}}$ is already multiplied by a factor of 10. At Uranus, the effect is less prominent, because of the smaller temperature jump implied by the lower nominal O/H: an increase of the opacities by a factor of 10 increases $y_{\mathrm{CO}}^{\mathrm{top}}$ by 25\%. It is thus important to see if we can set limits to this source of uncertainty.
    
    Interestingly, the formulation of \citet{Leconte2017} implies that the temperature jump cannot exceed a limit. This means that even if the opacities are significantly underestimated, the temperature jump cannot grow infinitely. This limit is given by the following $f_{\mu}$ factor:
    \begin{equation}
      f_{\mu} = \frac{1}{1+(\mu_{\mathrm{a}}/M_{\mathrm{H}_2\mathrm{O}} - 1) y_{\mathrm{H}_2\mathrm{O}}^{\mathrm{mass}}}
      \label{f_mu}
    \end{equation}
    and $f_{\mu}$ is a multiplicative factor that is applied to the temperature of a moist profile at the H$_2$O condensation level. In the previous expression, $\mu_{\mathrm{a}}$ is the mean molecular weight of the dry air and
    \begin{equation}
      y_{\mathrm{H}_2\mathrm{O}}^{\mathrm{mass}} = \frac{M_{\mathrm{H}_2\mathrm{O}}}{\mu}  y_{\mathrm{H}_2\mathrm{O}}^{\mathrm{bottom}}
    \end{equation}
    is the mass fraction of H$_2$O. Indeed, if we assume that a zone, in which dry convection would occur despite the presence of H$_2$O, can exist, then the virtual potential temperature of equation 8 in \citet{Leconte2017} should be approximately constant. If we now consider that the mole fraction of H$_2$O is negligible in the upper troposphere, the deep temperature cannot exceed the temperature given by the virtual potential temperature of the upper troposphere. This gives the limiting factor of equation~\ref{f_mu}. In practice, the temperature at the bottom of the atmosphere should be always lower than this limit, because of diffusive phenomena and moist convection that transport energy in the form of latent heat. We will explore the range of allowed thermal profiles by sampling the allowed profiles between the coldest one, which is given by the moist adiabat, and the warmest one, which is given by a moist adiabat in which the temperature at the H$_2$O condensation layer is multiplied by $f_{\mu}$. We emphasize that the 3-layer profiles always fall inside this range of profiles. In practice in this study, we compute 20 profiles from the coldest to the warmest, for each set of O/H, $y_{\mathrm{CH}_4}^{\mathrm{top}}$ and $K_{zz}$.

  \subsection{Step 2 - Calculating the thermochemical equilibrium}
  Once the internal elemental composition is estimated and the thermal profile derived, we can compute the atmospheric composition as a function of altitude assuming thermochemical equilibrium, i.e. in the absence of vertical mixing. This thermochemical equilibrium state is calculated by minimizing the Gibbs energy and these calculations are based on the algorithm of \citet{Gordon1994}. This equilibrium code has been developed by \citet{Agundez2014} and adapted for our purposes. The results of this code are then used as initial conditions of the thermochemical and diffusion model. This step enables us to speed up the latter computations by feeding the model with a non-empty atmosphere.

  \subsection{Step 3 - Calculating the thermochemistry and diffusion steady state}
  We use the thermochemical and diffusion model of \citet{Venot2012} adapted to the giant planet tropospheres to compute the steady state composition of the troposphere. This model has the advantage to be based on a chemical scheme that has been validated intensively by the combustion industry \citep{Bounaceur2007}. A full description of the model and its chemical scheme can be found in \citet{Venot2012} and can be retrieved from the KIDA database \citep{Wakelam2012}. 
  
  We start from an initial condition assuming chemical equilibrium as detailed in the previous section, assume a value for $K_{zz}$, and integrate the continuity equation over $\sim10^{8}$\,s usually. This enables reaching a steady state in all cases. As CO has also an external source in Uranus and Neptune \citep{Cavalie2014,Lellouch2005}, we have made sure that any reasonable change in the magnitude of the CO external source had no significant effect on the tropospheric CO mole fraction that is usually implicitly assimilated to an internal source. In Neptune, the external source is probably a relatively recent comet \citep{Lellouch2005,Luszcz-Cook2013} that should therefore not have had enough time to contaminate the upper tropospheric CO profile. In Uranus, the nature of the external source is still uncertain \citep{Cavalie2014}. However, the fact that the vertical mixing time through the stratosphere is more than two orders of magnitude higher than the integration time we need in our computations makes the tropospheric CO more sensitive to the internal source than to the external source. As a consequence, we can constrain the deep oxygen abundance with our model by fitting the observed upper tropospheric CO.

  \subsection{Step 4 - Tropospheric vertical mixing estimation and validation}
  In 1D thermochemical models, convective mixing is usually approximated by a vertical eddy diffusion coefficient $K_{zz}$. A thermochemical reaction can be quenched when the vertical diffusion timescale becomes shorter than the chemical timescale. Therefore, constraining the level at which the quenching of CO happens requires an estimate of vertical transport timescales in the tropospheres of giant planets. 
  
  As shown in previous studies, it appears that the final CO tropospheric mole fraction is sensitive to the K$_{zz}$ at the CO quench-level only. This quench-level remains typically around the same temperature level (i.e. $\sim$900\,K). So, we need to use a value relevant for this temperature level, and will assume uniform $K_{zz}$ coefficients. The validity of the latter assumption is further discussed in Section~\ref{insulation_layer}.
        
  Following the free-convection and mixing-length theories (MLT) of \citet{Stone1976} and \citet{Gierasch1985}, \citet{Visscher2010} give a generic form for $K_{zz}$:
  \begin{equation}
    K_{zz} \simeq \left(\frac{k_B F}{\rho m c_\mathrm{p}}\right)^{1/3}\times H.
    \label{Kzz_MLT}
  \end{equation} 
  This formulation, in which $k_B$ is the Boltzmann constant, requires the knowledge of the planet's internal heat flux $F$, the atmospheric mean mass density $\rho$, the atmospheric mean molecular mass $m$, the atmospheric mass specific heat at constant pressure $c_\mathrm{p}$, and the atmospheric scale height $H$. The value of $F$ has been measured by Voyager 2 (see Table~\ref{Obs_list}). This formulation is supposed to provide estimates valid within an order of magnitude. 
   
  The aforementioned formulation based on MLT thus depends on the composition (via $\rho$, $m$, $c_\mathrm{p}$, and $H$) and on the thermal structure (via $H$ and $c_\mathrm{p}$). The thermal structure in our models also depends on composition (via mean molecular weight effects and, to a lesser extent, $c_\mathrm{p}$). Therefore, we cannot compute $K_{zz}$ a priori and we will investigate a broad range of value (typically 5-6 orders of magnitude) in our thermochemical computations. Only after obtaining the thermal and composition results are we able to establish the validity range for $K_{zz}$ using the MLT formulation. 
  
  More recently, \citet{Wang2015} have used laboratory studies of turbulent rotating convection to derive a new formulation of $K_{zz}$ that provides estimates with a relative uncertainty of $\sim$25\%~only. We note that their formulation for low latitudes is similar to equation \ref{Kzz_MLT}, only corrected by a scaling factor. So, again, their formulation exhibits a dependence on temperature and composition.

\section{Observational constraints \label{Observation_constraints}}
  \subsection{Tropospheric abundances observed in giant planets}
  The thermochemical and diffusion model is initiated by specifying the deep abundances of the following elements : H, He, C, N, and O. We thus have to adjust the elemental abundances to ensure fitting the abundances of the main species in the upper troposphere, where their abundances have been measured. We have reviewed the abundances of species relevant to our model in all the Solar System giant planets and listed the values in \tab{Obs_list}. 
  
  The tropospheric CO mole fractions come from recent interferometric and space-based observations for Uranus and Neptune. On Uranus, \citet{Teanby2013} have obtained an upper limit of 2.1\dix{-9}, which is an improvement over \citet{Cavalie2008a} by almost an order of magnitude. The observations of \citet{Luszcz-Cook2013} that we use for Neptune have quite significantly revised the previously accepted value of 1.0\dix{-6} \citep{Rosenqvist1992,Marten1993,Marten2005,Lellouch2005,Lellouch2010,Fletcher2010}, and their result is confirmed (and the error bar is narrowed) by the recent IRAM-30m and Herschel/SPIRE observations (R. Moreno, priv. com.) that we use here as our nominal value: 0.2\dix{-6}. 
  
  The helium mole fraction we use for Uranus and Neptune come from the Voyager 2 measurement of \citet{Conrath1987} and from Infrared Space Observatory observations \citep{Burgdorf2003}. Both indicate that there is 15\% of helium in both atmospheres.
  
  Contrary to Jupiter and Saturn, Uranus and Neptune have cold enough tropopauses for CH$_4$ to condense and sharply decrease from their tropospheres to their stratospheres \citep{Lellouch2015}. \citet{Karkoschka2009} and \citet{Sromovsky2011,Sromovsky2014} for Uranus, and \citet{Karkoschka2011} for Neptune, have recently shown that the tropospheric CH$_4$ abundance is not uniform with latitude. The widely accepted measurements of \citet{Baines1995} ($\sim$2\% in both planets) seems to be representative of high latitudes, while the CH$_4$ equatorial tropospheric mole fraction is around 4\%. We will take this latter value as our nominal abundance for CH$_4$.

  \subsection{Upper tropospheric temperatures and internal heat fluxes}
  As a starting point for the temperature extrapolation in the troposphere, we take the temperature at 2\,bar in Uranus and Neptune, as measured by \citet{Orton2014a}, and \citet{Lindal1992}, respectively. 
  
  The computation of $K_{zz}$ requires the knowledge of the internal heat flux. We take the Voyager 2 measurements from \citet{Pearl1990} and \citet{Pearl1991} for Uranus and Neptune.

\begin{table*}
  \caption{Summary of observational data. }             
  \label{Obs_list}      
  \begin{center}          
  \begin{tabular}{cccccccc}
    \hline   
    Planet                             & Jupiter & Saturn & Uranus & Neptune \\
    \hline
    \y{He}$^{(1)}$                & 0.1359$\pm$0.0027       & 0.118$\pm$0.025       & 0.152$\pm$0.033      & 0.149$^{+0.017}_{-0.022}$ \\
    \y{CH}$_{4}$$^{(2)}$     & (2.04$\pm$0.50)\dix{-3} & (4.7$\pm$0.2)\dix{-3}  & 0.01-0.05                  & 0.01-0.05 \\
    \yco$^{(3)}$                    & (1.0$\pm$0.2)\dix{-9}     & $<$10$^{-9}$               & $<$2.1\dix{-9}             & (0.20$\pm$0.05)\dix{-6} \\
    $T^{(4)}$                        & 207.1                                 &  132.9                             & 102.9                               & 93.1 \\
    $F^{(5)}$                         & 5.44$\pm$0.43                & 2.01$\pm$0.14            & 0.042$\pm$0.047       & 0.433$\pm$0.046 \\
    $T_\mathrm{int}^{(6)}$    & 99.0                               & 77.2                            & 29.3                            & 52.6   \\
    \hline  
  \end{tabular}
  \end{center}
  \scriptsize{\underline{Notes:}} \y{i} are mole fractions, and F is the internal heat flux in W$\cdot$m$^{-2}$, temperatures are in K. \\
  \scriptsize{\underline{References:}} $^{(1)}$ \citet{vonZahn1998} and \citet{Niemann1998} for Jupiter, \citet{Conrath2000} for Saturn, \citet{Conrath1987} for Uranus, and \citet{Burgdorf2003} for Neptune. \\
  $^{(2)}$ \citet{Wong2004} for Jupiter, \citet{Fletcher2009,Fletcher2012} for Saturn, \citet{Lindal1987,Baines1995,Karkoschka2009,Sromovsky2014} for Uranus, and \citet{Lindal1990,Baines1995,Karkoschka2011} for Neptune. The range for Uranus and Neptune represents the observed latitudinal variability. \\
  $^{(3)}$  \citet{Bezard2002} for Jupiter, \citet{Cavalie2009} for Saturn, \citet{Teanby2013} for Uranus, and \citet{Luszcz-Cook2013} and \citet{Moreno2011} for Neptune. \\
  $^{(4)}$ Galileo measurement of \citet{Seiff1998} for Jupiter, equatorial average at 1\,bar obtained with Cassini/CIRS by \citet{Fletcher2016} for Saturn, Spitzer measurements from \citet{Orton2014a} for Uranus, Voyager 2 occultation observations from \citet{Lindal1992} for Neptune. Temperatures at 2\,bar, except for Saturn (1\,bar). \\
  $^{(5)}$ Voyager 2 measurements from \citet{Hanel1981} for Jupiter, \citet{Hanel1983} for Saturn, \citet{Pearl1990} for Uranus, and \citet{Pearl1991} for Neptune.  \\
  $^{(6)}$ $T_\mathrm{int}$ is computed from $F=\sigma T_\mathrm{int}^4$.
\end{table*}

\section{Results \label{Results}}
Hereafter, we present the results of our exploration of a 4D parameter space (O/H, C/H, temperature, $K_{zz}$) in thermochemical and diffusion computations in an attempt to constrain the deep oxygen abundance of Uranus and Neptune by fitting the observed upper tropospheric CO mole fraction. From this set of results, we also show how the O/H ratios derived from a thermochemical model that considers ``3-layer'' thermal profiles as nominal compare to previously published results in which dry or wet adiabats had been assumed. We want to keep the reader aware of the fact that the ``3-layer'' thermal profiles have been shown to be stable in 1D \citep{Leconte2017}, but may not be in a 3D treatment. The model is affected by sources of uncertainties other than those investigated in this 4D study and they are described in Section~\ref{Limitations}. So, in the following sections, we present the results of our 4D grid computations, and we provide the reader with the deep O/H obtained for Uranus and Neptune with our model, when assuming our nominal input parameters.

In the text that follows, the O/H and C/H ratios will be presented as a function of the O and C solar abundances, i.e. 8.73\,dex for O and 8.39\,dex, as reported by \citet{Lodders2010}.

  \subsection{$K_{zz}$ a posteriori validation}
  The estimation of the deep composition at step 0 and the calculation of the thermal profile at step 1 of our modeling enables us to compute estimates of $K_{zz}$ using the MLT or the \citet{Wang2015} formulation. For simplicity, we take the MLT formulation. The results presented below include an a posteriori consistency check between the estimated $K_{zz}$ and the value assumed in the thermochemical modeling (step 3). This is the step 4 of our analysis. For Uranus and Neptune, we find $K_{zz}\simeq10^8$\,\Kunit.

  \subsection{Exploring the 4D parameter space}
    \subsubsection{Methodology}
    The $y_{\mathrm{CO}}^{\mathrm{top}}$ we obtain from our computations depends on the following quantities: the deep O abundance, $y_{\mathrm{CH}_4}^{\mathrm{top}}$ (as it controls $X_C$), the deep tropospheric temperature, and $K_{zz}$. To properly explore this 4D parameter space, we have run our models detailed in steps 0-3 of Section~\ref{Models} in a loop format, by varying these quantities. 
  
    For the deep oxygen abundance we go from 10 to 300$\Sun$ for Uranus and from 10 to 600$\Sun$ for Neptune, and for the $y_{\mathrm{CH}_4}^{\mathrm{top}}$ mole fraction we go from 1.5\% to 4.5\%. The temperature profiles are computed once these quantities are fixed.
  
    The link between the upper tropospheric CO mole fraction and the deep O abundance depends on the level at which the quenching of CO occurs, i.e. the level at which the chemical lifetime of CO, which is mainly governed by temperature, equals its diffusion timescale given by $K_{zz}$. This level is usually located below the H$_2$O cloud base, i.e. in a region where a dry adiabat prevails in any case. Because the behavior of the thermal profile remains poorly constrained around the H$_2$O condensation level, and because the thermal profile should follow a simple dry adiabat below this level, we can explore the various possible deep thermal profiles by starting a dry adiabatic extrapolation from below the H$_2$O cloud base. The starting temperature range from the temperature given by the moist adiabat, i.e. the coldest case, and the extreme case derived from \citet{Leconte2017}. The 3-layer profiles fall, by definition, in this range. When computing the extreme thermal profiles, we apply the $f_{\mu}$ factor from equation \ref{f_mu} to the temperature at the level where the temperature jump occurs in the 3-layer profiles. We thus have the possibility to limit the range of possible temperature jumps, from a situation in which the opacities would be so low that no jump would be formed and the temperature profile would finally follow the moist adiabat to a situation in which the opacities would be high enough so that the temperature jump would reach its allowed limit. We sample this range with 20 profiles.
  
    Finally, we vary the values of $K_{zz}$ over 5-6 orders of magnitude around the value that provides a fit to $y_{\mathrm{CO}}^{\mathrm{top}}$ in the ``3-layer'' case.

    \subsubsection{Effect of $K_{zz}$}  
\begin{figure}[htp]
  \begin{center}
  \includegraphics[width=10cm,keepaspectratio]{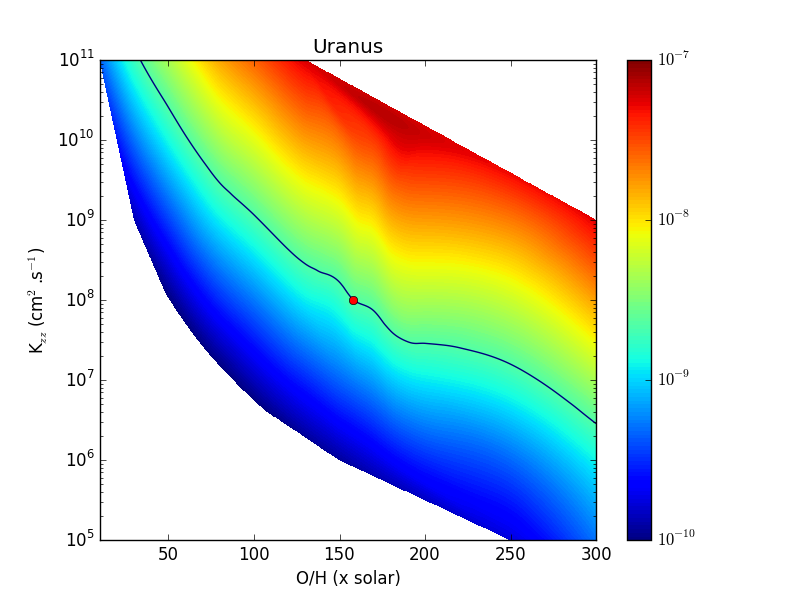}
  \caption{Relationship between the magnitude of $K_{zz}$ in the troposphere, the deep O abundance (expressed in solar abundances), and the resulting upper tropospheric CO for Uranus. In these simulations, we use ``3-layer'' thermal profiles and $y_{\mathrm{CH}_4}^{\mathrm{top}}$ is set to 0.04. The black curve represents the values of $K_{zz}$ and O/H that fit the observations of $y_{\mathrm{CO}}^{\mathrm{top}}$. The red dot is the nominal case ($y_{\mathrm{CH}_4}^{\mathrm{top}}=$0.04, $K_{zz}=10^8$\,\Kunit, ``3-layer'' profile). At Uranus, this curve is an upper limit for O/H. We refer the reader to the online version of the paper for the color scale.}
  \label{Kzz_O_Ura}
  \end{center}
\end{figure}

    This case is obvious as the deep O abundance required to fit the upper tropospheric CO decreases with increasing $K_{zz}$. The relationships for Uranus and Neptune are presented in \figs{Kzz_O_Ura}{Kzz_O_Nep}. They show the range of values for ($K_{zz}$,O/H) that are compliant with observations. We remind the reader that the observations at Uranus have only enabled setting an upper limit on $y_{\mathrm{CO}}^{\mathrm{top}}$, and all values of O/H below the black curve shown in the figure are thus not ruled out.

\begin{figure}[htp]
  \begin{center}
  \includegraphics[width=10cm,keepaspectratio]{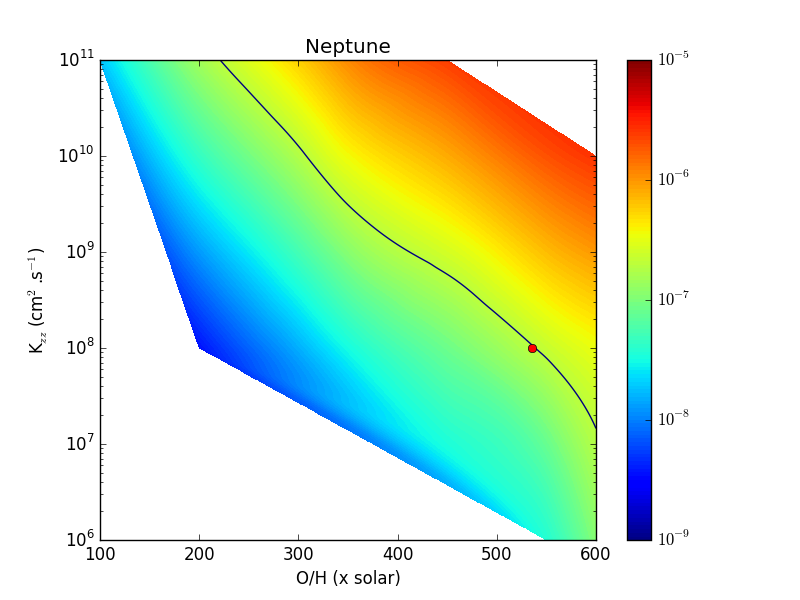}
  \caption{Same as \fig{Kzz_O_Ura} for Neptune.}
  \label{Kzz_O_Nep}
  \end{center}
\end{figure}

    \subsubsection{Effect of $y_{\mathrm{CH}_4}^{\mathrm{top}}$}
    \citet{Cavalie2014} investigated the effect of the tropospheric CH$_4$ mole fraction on the deep O abundance in Uranus. They found that an increase of the CH$_4$ abundance implied a decrease of the required O abundance to fit the tropospheric CO, essentially because increasing the C abundance implies an increase C atoms available to form CO. However, they used strictly similar thermal profiles for their two cases and neglected the influence of CH$_4$ on the thermal profile. 
      
    In this work, we take into account the influence of the mean molecular weight $\mu$ and of the $c_p$ of CH$_4$ and H$_2$O when computing the thermal profile. Increasing the C abundance now also implies increasing $\mu_a$ in equation \ref{Ledoux_criterion} and thus having a temperature jump located deeper in the troposphere. Moreover, the thermal gradient is lower because of a higher $\mu c_p$. Consequently, the deep tropospheric temperatures are colder when $y_{\mathrm{CH}_4}^{\mathrm{top}}$ increases, if other quantities are kept constant. However, the increase of available C atoms to form CO remains the dominant effect when $y_{\mathrm{CH}_4}^{\mathrm{top}}$ (and thus $X_C$) increases. So, if $y_{\mathrm{CH}_4}^{\mathrm{top}}$ (and thus the deep C abundance) increases, a lower deep O abundance is needed to fit the tropospheric CO.

    \subsubsection{Effect of the temperature jump magnitude in the H$_2$O condensation region \label{Temp_jump}}
    We have investigated 20 profiles for each set of O/H, C/H and $K_{zz}$ that range from the coldest possible profile, i.e. a wet adiabat, to the warmest one, i.e. a wet adiabat in which the multiplication factor $f_{\mu}$ is applied to the temperature in the layer where H$_2$O condenses. The latter is the extreme profile of the ``3-layer'' case and the multiplication factor is given by equation\,\ref{f_mu}. The nominal 3-layer profiles always fall within this range.
    
\begin{figure}[htp]
  \begin{center}
  \includegraphics[width=6cm,keepaspectratio]{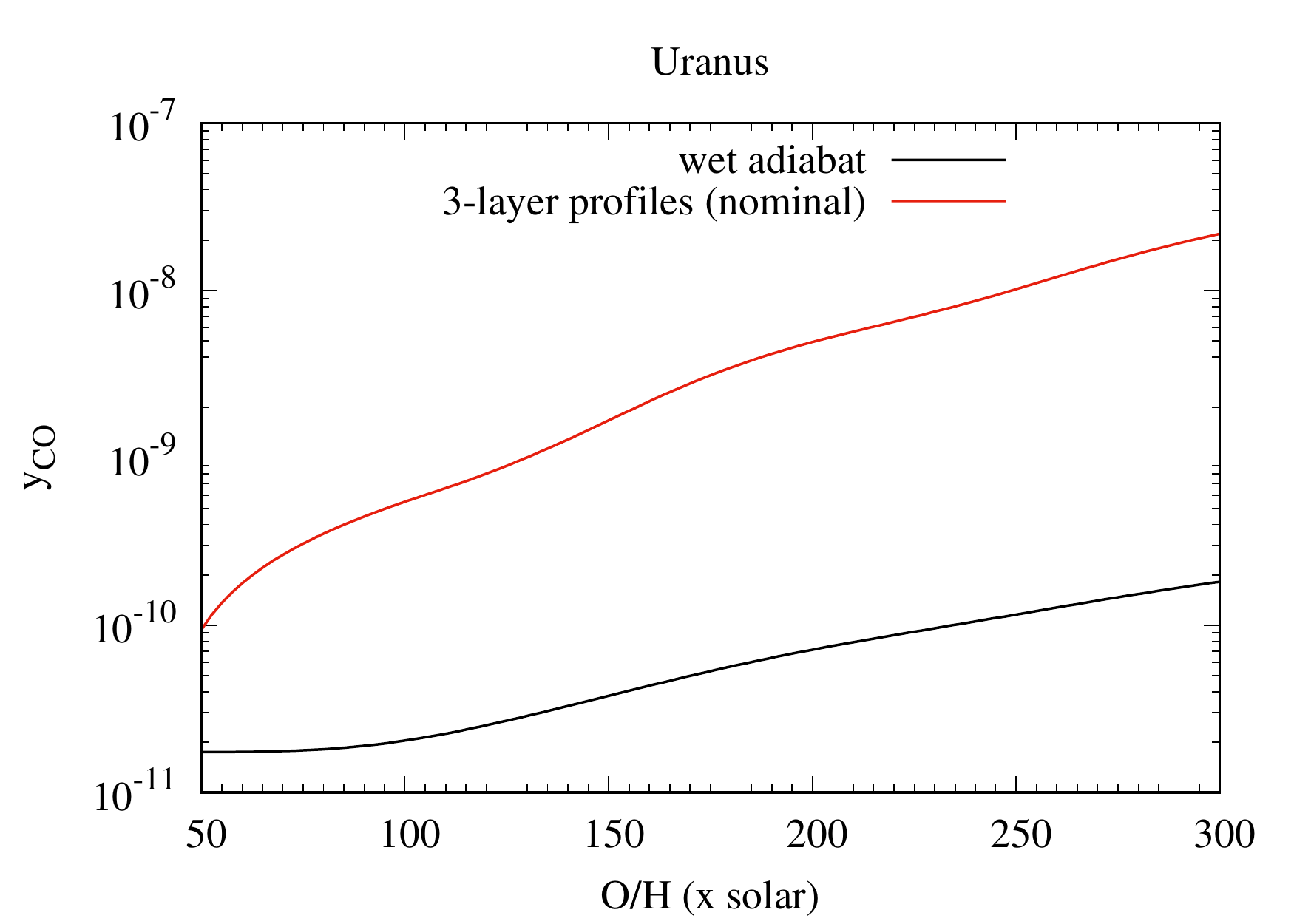}
  \includegraphics[width=6cm,keepaspectratio]{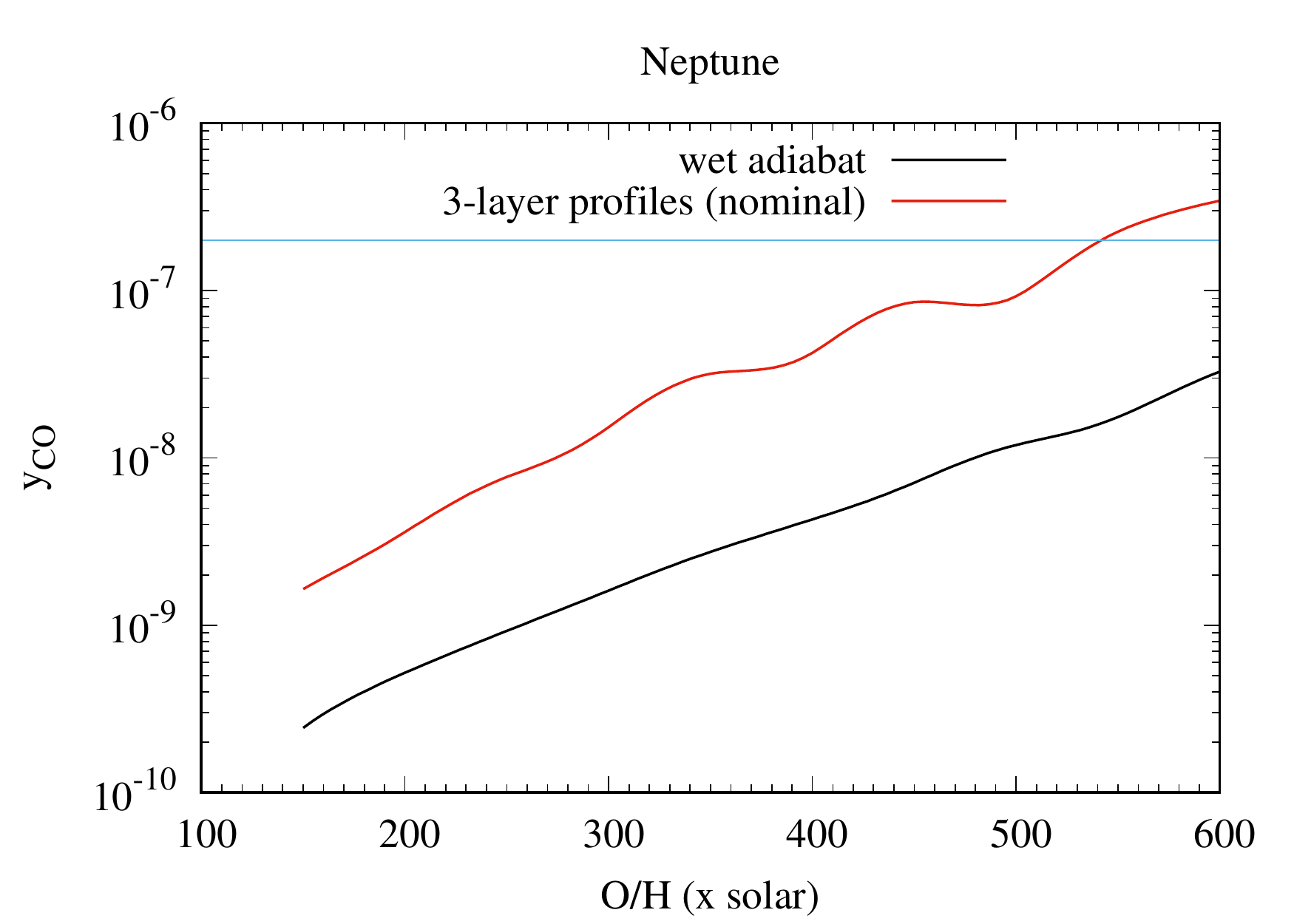}
  \caption{Comparison between the upper tropospheric CO mole fraction produced with wet adiabats and nominal 3-layer profiles as a function of the deep O/H ratio, for Uranus and Neptune. The latter produce significantly more CO for a given O/H ratio. We use the nominal values for $y_{\mathrm{CH}_4}^{\mathrm{top}}$ (0.04) and $K_{zz}$ ($10^8$\Kunit). The $y_{\mathrm{CO}}$ upper limit for Uranus and measurement in Neptune are indicated with a thin blue line. }
  \label{wet_3_layer}
  \end{center}
\end{figure}

\begin{figure*}[htp]
  \begin{center}
  \includegraphics[width=10cm,keepaspectratio]{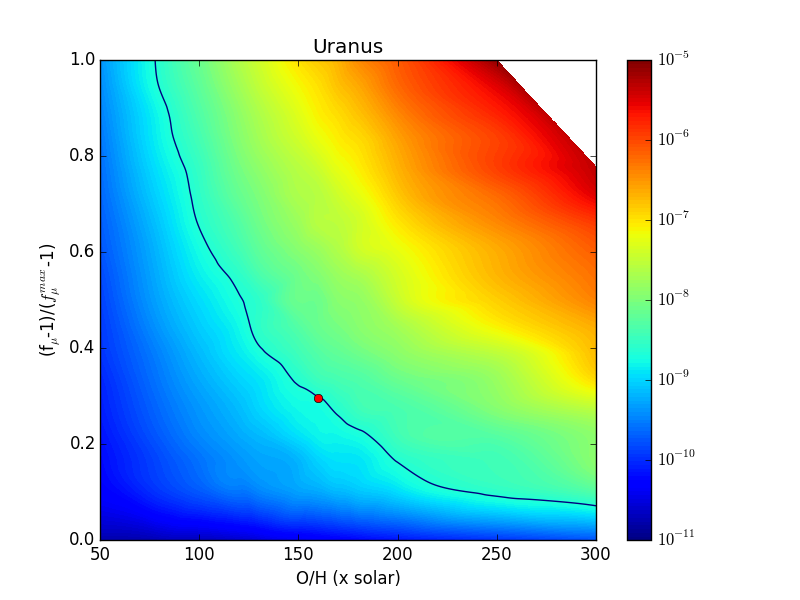}
  \caption{Relationship between the multiplication factor applied to the temperature at the condensation level of H$_2$O (normalized to its maximal allowed value, see equation\,\ref{f_mu}), the deep O abundance, and the resulting upper tropospheric CO for Uranus. We use the nominal values for $y_{\mathrm{CH}_4}^{\mathrm{top}}$ (0.04) and $K_{zz}$ ($10^8$\Kunit). The black curve represents the values that fit the observations of $y_{\mathrm{CO}}^{\mathrm{top}}$. The red dot is the nominal case ($y_{\mathrm{CH}_4}^{\mathrm{top}}=$0.04, $K_{zz}=10^8$\,\Kunit, ``3-layer'' profile). At Uranus, this curve is an upper limit for O/H. We refer the reader to the online version of the paper for the color scale.}
  \label{temp_O_Ura}
  \end{center}
\end{figure*}

    The results are quite difficult to present, in the sense that each individual O/H results in a different $f_{\mu}$ factor. A first indication can be obtained by comparing the results obtained with wet adiabats and nominal 3-layer profiles. \fig{wet_3_layer} essentially shows that the 3-layer profiles require significantly lower O/H ratios to produce a given upper tropospheric CO mole fraction. To illustrate the full range of 3-layer profiles that are allowed and their impact on the results, we choose to normalize the results with respect to the maximal magnitude of multiplication factor for each O/H, noted $f_{\mu}^{\mathrm{max}}$. We thus compute the curves for $f_{\mu}-1$$/$$f_{\mu}^{\mathrm{max}}-1$ as a function of O/H. The values range from the wet adiabat, for which $f_{\mu}$$=$1 and $f_{\mu}-1$$/$$f_{\mu}^{\mathrm{max}}-1$=0, to the maximum amplitude temperature jump, for which $f_{\mu}$$=$$f_{\mu}^{\mathrm{max}}$ and $f_{\mu}-1$$/$$f_{\mu}^{\mathrm{max}}-1$=1. \figs{temp_O_Ura}{temp_O_Nep} display the results for both planets. In this case, $K_{zz}$ is fixed to 10$^8$\,\Kunit. The downside is that the absolute magnitude of $f_{\mu}$ is not directly accessible to the reader. It essentially shows that a much higher O/H would be needed to fit the data if a pure wet adiabat was used in Uranus, and that even the warmest profile cannot produce enough CO in the upper troposphere if the deep O abundance is lower than $\sim$80 times solar. However, the tropospheric CO value is an upper limit in Uranus and this constraint is therefore useless. In Neptune, the fraction of the maximal multiplication factor sufficient to produce enough CO in the troposphere decreases quickly with increasing O/H, as this multiplication factor is already huge for O/H of several hundreds and a small fraction of it produces a big temperature jump in the zone where H$_2$O condenses. The results also show that a minimum of O/H$\sim$190 times solar seems to be a lower limit in Neptune with our model.

\begin{figure*}[htp]
  \begin{center}
  \includegraphics[width=10cm,keepaspectratio]{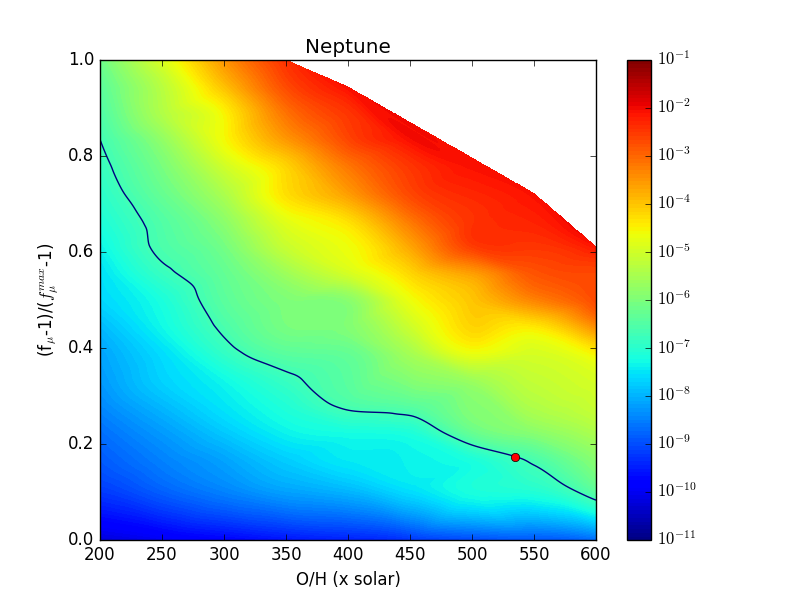}
  \caption{Same as \fig{temp_O_Ura} for Neptune.}
  \label{temp_O_Nep}
  \end{center}
\end{figure*}

  \subsection{Uranus nominal case \label{sub:Uranus}}
\begin{figure*}[htp]
  \begin{center}
  \includegraphics[width=10cm,keepaspectratio]{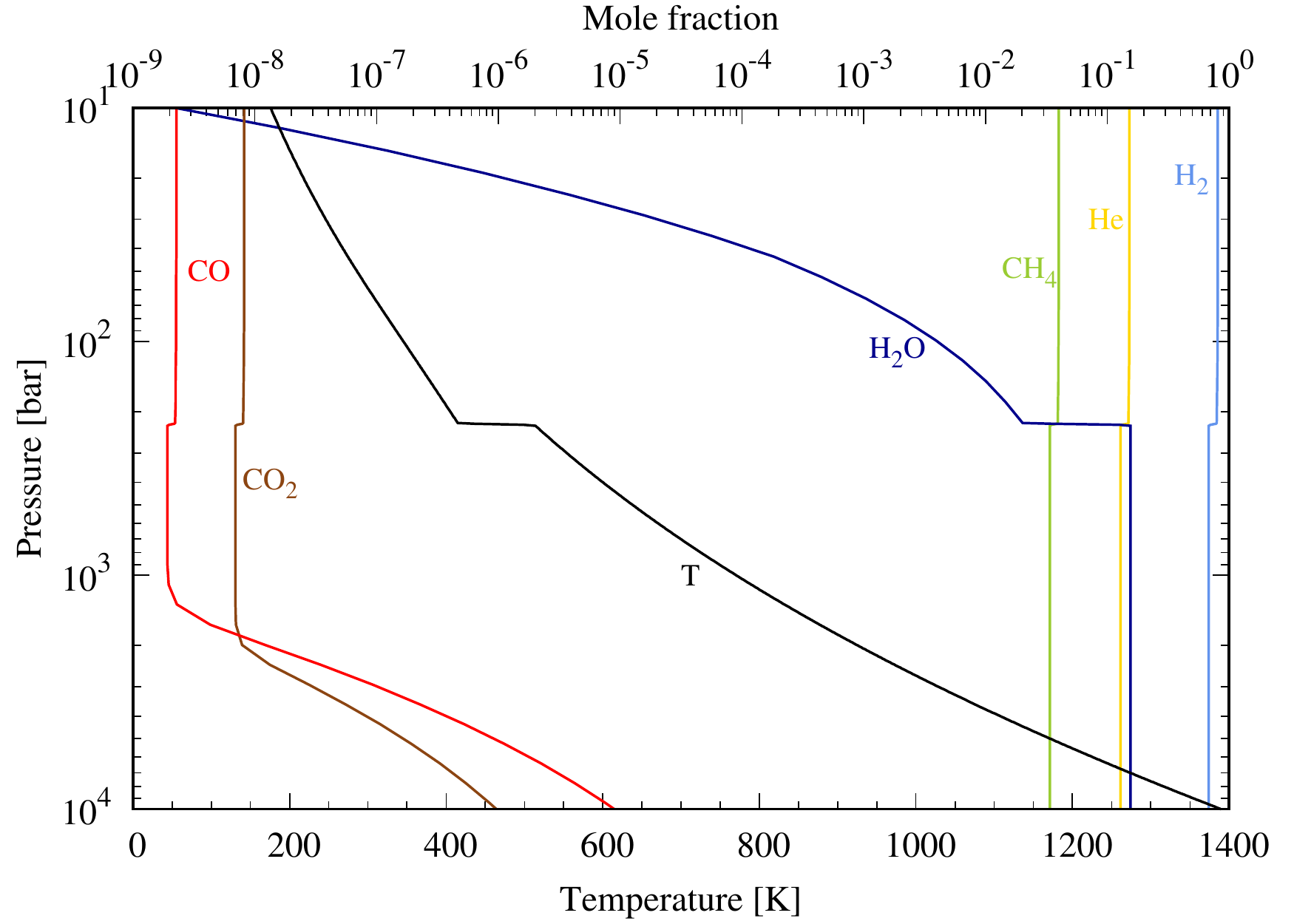}
  \caption{Molar fraction profiles in the troposphere of Uranus obtained with the thermochemical and diffusion model, targeting the upper limit of 2.1\dix{-9} upper tropospheric CO mole fraction, as determined by \citet{Teanby2013}. The deep O/H ratio is 160 times the solar value. We use the nominal values for $y_{\mathrm{CH}_4}^{\mathrm{top}}$ (0.04) and $K_{zz}$ ($10^8$\Kunit). The ``3-layer'' temperature profile in the troposphere corresponding to this O abundance and with which these abundance profiles have been obtained is shown with a black solid line. CO is quenched between 1 and 2\,kbar, at $T$$=$900-1000\,K.}
  \label{resultats_Ura}
  \end{center}
\end{figure*}

  ``3-layer'' thermal profiles of Uranus present a less spectacular transition zone than in Neptune, but still modify significantly the upper limit on the O/H ratio when compared to cases in which we would use dry or wet adiabats. We find that reproducing the observed upper limit of 2.1\dix{-9} of CO and the reported He and CH$_4$ abundances (see \tab{Obs_list}) requires C/H and O/H ratios of 75, and 160 times the solar value. The resulting abundance profiles are displayed in \fig{resultats_Ura}. As in the Neptune case, using dry or wet adiabats would lead to much lower amounts of CO when starting from the same elementary abundances.

  \subsection{Neptune nominal case \label{sub:Neptune}}  
  As stated above, Neptune is expected to be the most ``symptomatic'' case demonstrating the impact of the mean molecular weight gradient in the transition zone from the water-rich to the water-poor region in its troposphere. It leads to a ``3-layer'' thermal profile with a significant departure from the purely dry or wet adiabats that had been used previously in the literature (e.g., \citealt{Lodders1994,Luszcz-Cook2013}). To fit the nominal 0.2\dix{-6} of tropospheric CO (see \tab{Obs_list}), we had to set the O/H ratio to $\sim$540 times the solar value. The corresponding ``3-layer'' thermal profile is displayed in \fig{resultats_Nep}. The nominal C/H ratio is 40 times the solar value. This value is rather low, if compared to a value that would simply be deduced from the upper atmospheric abundances of CH$_4$. The difference is caused by the high internal abundance of O and thus the high deep H$_2$O abundance. Indeed, we find in this calculation that H$_2$O is more abundant than H$_2$ in Neptune's interior.

\begin{figure*}[htp]
  \begin{center}
  \includegraphics[width=10cm,keepaspectratio]{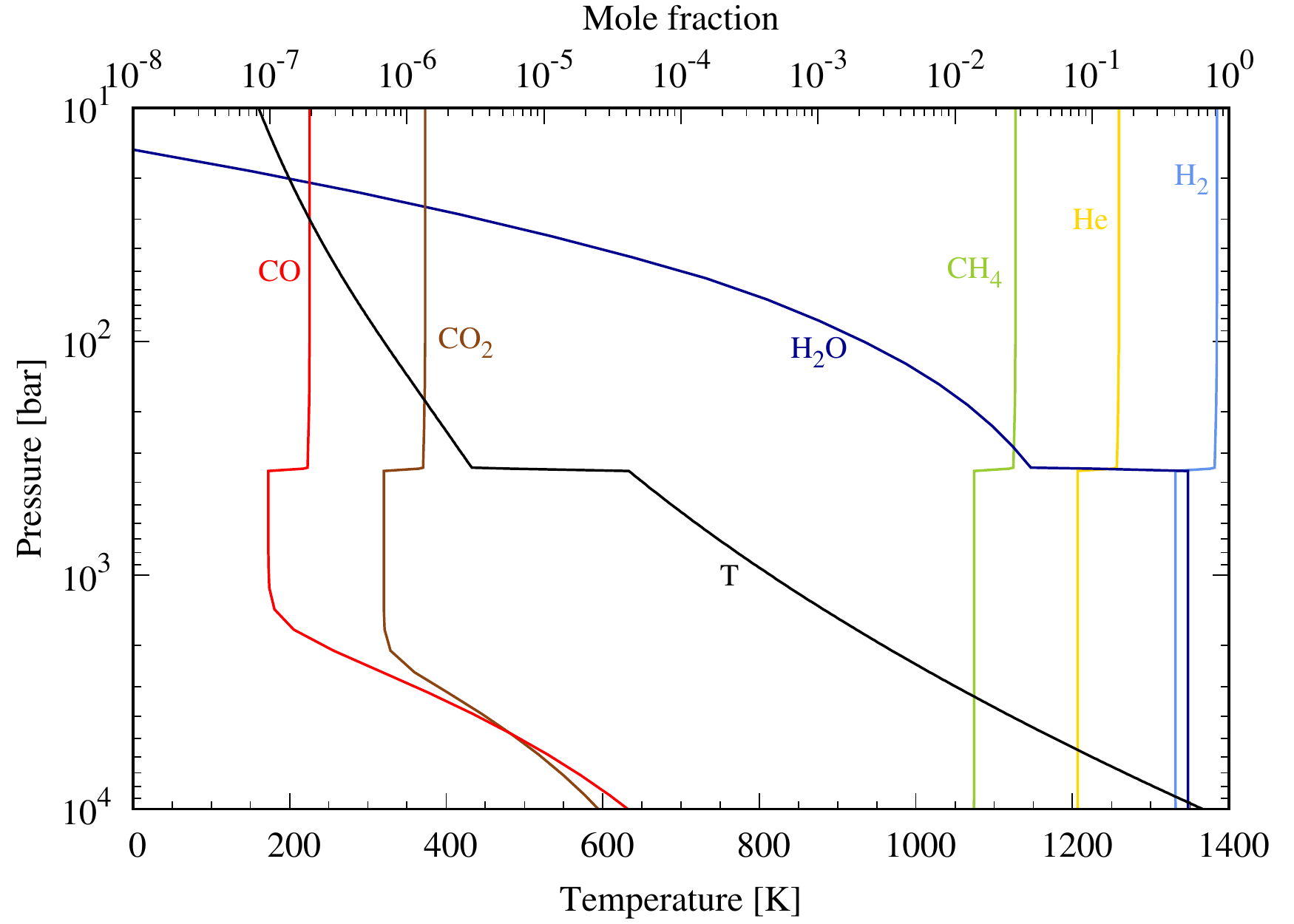}
  \caption{Molar fraction profiles in the troposphere of Neptune obtained with the thermochemical and diffusion model, targeting a 0.2\dix{-6} upper tropospheric CO mole fraction. The deep O/H ratio is 480 times the solar value. We use nominal the values for $y_{\mathrm{CH}_4}^{\mathrm{top}}$ (0.04) and $K_{zz}$ ($10^8$\Kunit). The ``3-layer'' temperature profile in the troposphere corresponding to this O abundance and with which these abundance profiles have been obtained is shown with a black solid line. CO is quenched between 1 and 2\,kbar, at $T$$=$800-900\,K. The increase in mole fraction of compounds other than H$_2$O between 300 and 400\,bar is caused by the condensation of H$_2$O.}
  \label{resultats_Nep}
  \end{center}
\end{figure*}

\section{Discussion \label{Discussion}}
This section will be divided in two parts. In the first one, we will discuss the results of the nominal model, as presented in Section~\ref{Results}. In the second part, we will discuss the microwave spectral counterpart of our model results to further constrain their validity.

  \subsection{Nominal results}
  We find nominally that Uranus has a C/H ratio of 75 times the solar value and a O/H ratio below 160 times the solar value. For Neptune, we get a C/H 40 times the solar value and a O/H 540 times the solar value.
  
    \subsubsection{A deep compositional difference between Uranus and Neptune? \label{composition}}
    One of the striking compositional differences in Uranus and Neptune atmospheres is their CO abundances, while their He and CH$_4$ abundances and D/H ratios are relatively similar \citep{Baines1995,Karkoschka2009,Karkoschka2011,Sromovsky2011,Sromovsky2014,Feuchtgruber2013}. Indeed, they differ by at least two orders of magnitude in the troposphere \citep{Teanby2013,Luszcz-Cook2013}. However, previous evaluations of the O/H ratios in these bodies using the quench-level approximation and dry/wet adiabats never differed by more than a factor of two \citep{Lodders1994,Luszcz-Cook2013,Cavalie2014}. Accounting for the stability criterion of \citet{Leconte2017} in the water condensation layers has a significant effect on the planets thermal profile and hence on the derived O/H values. Our nominal values show that O/H$<$160 and O/H$=$540 times the solar value for Uranus and Neptune, respectively. So, there is a factor of $\geq$3.3 difference between these two values. The two-order of magnitude difference between the upper tropospheric CO mole fraction on Uranus and Neptune is still caused by a much smaller difference between their deep H$_2$O abundances. This can be understood from the equilibrium equation \ref{equilibrium_reaction}. An increase on H$_2$O implies a linear decrease of H$_2$. As the power on H$_2$ is 3  (and the CH$_4$ abundance is relatively constant), CO has to increase non-linearly to equilibrate the reaction. This effect is shown for both planets on \fig{CO_H2O}, where we plot the relationship between the CO and H2O abundances at the quench level for our nominal cases for $K_{zz}$ and $y_{\mathrm{CH}_4}^{\mathrm{top}}$.  Another interesting feature of our results is that a similar CH$_4$ abundance in Uranus and Neptune does not reflect in a similar C/H ratio. This can be understood from equations \ref{eq_X_He} and \ref{eq_X_C}. Indeed, the very high O/H ratio of Neptune implies a lower C/H ratio than in Uranus to reproduce both the CO and CH$_4$ abundances in the upper troposphere. In Uranus, the deep H$_2$O abundance is comparable to the He abundance, while the deep H$_2$O abundance even exceeds the H$_2$ abundance in Neptune.

 \begin{figure}[htp]
  \begin{center}
  \includegraphics[width=10cm,keepaspectratio]{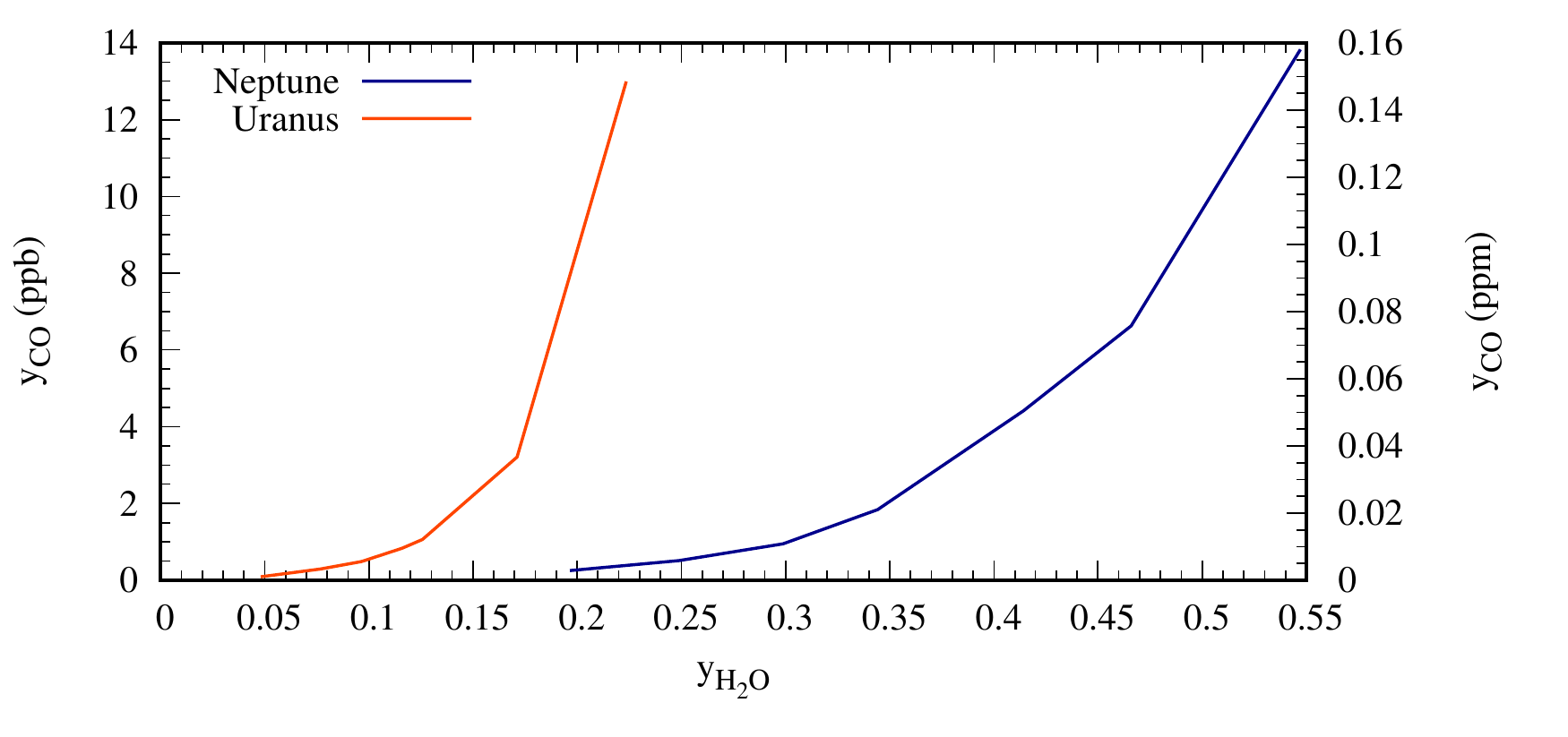}
  \caption{Relationship between the deep H$_2$O abundance and the CO abundance at the quench level in Uranus and Neptune, taking nominal conditions for $K_{zz}$ and $y_{\mathrm{CH}_4}^{\mathrm{top}}$. CO increases non-linearly with respect to H$_2$O, to compensate for the decrease of H$_2$ which has a stoichiometric coefficient of 3 in the equilibrium reaction.}
  \label{resultats_Nep}
  \end{center}
\end{figure}

    We confirm that a large difference in terms of CO abundance (two orders of magnitude) does not imply such a large difference in terms of O/H ratios in the two planets in our nominal model. This rather small difference seen in their O/H ratios, compared to their CO abundances, is then mostly caused by the 10\,K difference at 2\,bar, where we initiate our thermal profiles.     
    
    \subsubsection{Consistency with interior and formation models \label{consistency}}
    While the tropospheric mole fractions of CO are very different in Uranus and Neptune, their D/H ratios are quite similar, according to \citet{Feuchtgruber2013}. One of the possible explanations presented by these authors, based on the interior models of \citet{Helled2011} and \citet{Nettelmann2013}, is that the cores of the planets may be more rocky than icy, with rock fraction of 68-86\%, for their D/H to be still representative of the one seen in Oort cloud comets. The O/H in core ices implied in such cases ranges from 79 to 172 times the solar value for Uranus and from 68 to 148 times the solar value for Neptune \citep{Feuchtgruber2013}. These numbers, relevant only if the planets went through complete mixing at least once in their histories, are compatible with the O/H in the outer envelope of Uranus but seem nevertheless too low for Neptune's core ices to result in a sufficiently high outer envelope O/H ratio compared to our nominal results.
  
    A scenario has been proposed recently to reconcile the D/H measurements at Uranus and Neptune with the Oort cloud comet value by \citet{Ali-Dib2014}. They propose that Uranus and Neptune formed on the CO snowline and that their ices were mostly composed of CO ices rather than H$_2$O ices. Therefore, the present day atmospheric D/H would be representative of the dilution of the small fraction of D-enriched Oort-comet-like H$_2$O ices in the H$_2$O coming from the thermochemical conversion of the more abundant primordial CO. This scenario then predicts that oxygen should be enriched in a similar way as carbon in both planets. Our simulation results show that this is probably neither the case for Uranus nor for Neptune and seem to contradict the model of \citet{Ali-Dib2014}. We find C/O ratios of $>$0.21 and $\sim$0.03, respectively, for the two planets. In Neptune, the C/O ratio could be brought back to one if CO tropospheric mole fraction was several orders of magnitude below its measured value, i.e., at the lower end of the measurement range of \citealt{Luszcz-Cook2013}). However, recent Herschel-SPIRE and IRAM-30m observations show that the CO tropospheric mole fraction is rather in the upper part of the measurement range of \citet{Luszcz-Cook2013}, with a measured value of \yco~of 0.20$\pm$0.05 ppm (R. Moreno, priv. com.). In any case, the model of \citet{Ali-Dib2014} seems not to be compatible with the Nice model \citep{Gomes2005,Morbidelli2005,Tsiganis2005} and it would anyway need to be improved by accounting for the protoplanetary disk temporal evolution as well as ice and core formation kinetics to better constrain the validity of their assumption that core formation on the CO snowline is possible.
        
    Our results can further be used for direct comparison with formation model predictions. For instance, \citet{Hersant2004} predicted the O/H enrichment in Uranus and Neptune by exploring the range of possible C/H values when assuming that heavy elements were trapped by clathration. For instance, if C was carried by only CO during the formation of planetesimals, then 5.75 H$_2$O molecules were required to trap CO in clathrates. Then, the resulting C/O ratio is 1/(1$+$5.75)$=$0.15. However, their prediction is less extreme and is as follows: for C/H ratios of 40, 60, and 80 times the solar value, the O/H ratio should be 87, 130, and 176 times the solar value (based on solar abundances of \citealt{Anders1989}). In this case, half of the carbon comes from clathrated CO and half from pure CO$_2$ ice, and the resulting C/O ratio is 1/(2$\times$1/2 + 5.75$\times$1/2 + 1/2)$=$1/4.375$=$0.23. For reference, the solar C/O ratio is 0.46, according to \citet{Lodders2010}. The lower limit we obtain on the C/O ratio of Uranus is at the limit of compatibility with the case in which carbon equally comes from clathrated CO and CO$_2$ ice. On the other hand, the Neptune C/O ratio we derive would become compatible with the case in which all carbon comes from clathrated CO, as long as the efficiency of clathration was about 20\%. 
    
    The rather large difference in terms of C/O ratio in Uranus and Neptune seems to imply different condensation environments, which may be difficult to explain. The question then is, whether the Uranus upper tropospheric CO value is representative for the CO quench level value? Actually, convection could be much less efficient than assumed in our computations. We have indeed based our computations of the tropospheric $K_{zz}$ of Uranus on the nominal value of $F$ reported in \citet{Pearl1990}. However, the error bars on $F$ are such that even a zero value is compatible with the Voyager observations. A lower $F$ implies a lower $K_{zz}$, as $K_{zz}\propto F^{1/3}$ (see equation \ref{Kzz_MLT}). We can then reversely use our model to constrain $K_{zz}$ to bring the O/H of Uranus to an O/H compatible with the clathration scenario of \citet{Hersant2004}. If the O/H of Uranus is set to $\sim$450 times the solar value, so that C/O$\sim$0.15, then C/H is $\sim$50 times solar and $K_{zz}$$\sim$10$^5$\Kunit. Bringing the Uranus O/H further up to the Neptune nominal value implies $K_{zz}$ also $\lesssim$10$^5$\Kunit. It is thus possible that Uranus and Neptune are more similar regarding their oxygen reservoir than their tropospheric CO seem to tell us, provided that convection is strongly inhibited in the interior of Uranus. This seems to be supported by recent internal structure and evolution modeling. \citet{Nettelmann2016} attempt to fit both the low luminosity of Uranus and its gravitational data with an internal structure and evolution model. Achieving this requires a thermal boundary around 0.1\,Mbar that breaks the adiabat. They also find that fitting $J_4$ for Uranus requires an O/H of less than 30, while C/H is about 80. However, they do not consider C/O$>$1 to be a good assumption for the core, implying that deep stratification is the cause for a lower O abundance at observable levels.

    We can also compute the outer envelope heavy element mass fraction from the deep abundances of the species accounted for in the model and check whether they are in agreement with the internal structures inferred by \citet{Nettelmann2013}. For Uranus and Neptune, we find $Z_1$$<$55\%~and $=$80\%, respectively. The Neptune value is above the limit of $\sim$65\% allowed by their model (see figure 3 in \citealt{Nettelmann2013}). Only if we consider the most extreme case of temperature jump in our simulations, in which the O/H of Neptune cannot be lower than 190 times solar, can we get within the limit of \citet{Nettelmann2013}, with $Z_1$ equal to 58\%. Our Uranus O/H value, being an upper limit higher than the range of values allowed by their model, is not constraining. Figure 3 of \citet{Nettelmann2013} indicates an upper limit for $Z_1$ in Uranus of $\sim$10\%, which in turn would imply an O/H ratio much lower than our upper limit. Carbon which is mainly carried by CH$_4$ already represents a mass fraction $\geq$10\% in the outer envelope, and a lower O/H would result in a higher C/H and C mass fraction. Thus, it seems difficult to reconcile the model of \citet{Nettelmann2013} of Uranus with upper tropospheric measurements and their implication on deep tropospheric elemental composition. The case of Uranus thus remains puzzling, and more generally, the formation of Uranus and Neptune is difficult to explain \citep{Helled2014}. It thus underlines the need for new data. In this sense,  dedicated orbital missions with  atmospheric descent probes are highly desirable \citep{Arridge2012,Arridge2014,Masters2014,Turrini2014}. 
    
    Finally and as will be discussed in Section~\ref{Chemical_scheme}, our model may underestimate the production of CO. In this case, the model would require lower O abundances and bring the results closer to an agreement with the predictions from \citet{Nettelmann2013} for both planets.

  \subsection{Radio spectrum of Uranus and Neptune}
  To further constrain the validity of the results of this paper, we present radiative transfer simulations in the microwave range with the temperature and abundance profiles of our thermochemical computations. It is indeed worth checking whether the temperature jump implied for high deep O abundances and the sharp composition transition in the H$_2$O condensation zone translate into spectra that are in agreement with observations in the mm-cm range (e.g. \citealt{dePater1991}).
  
  We thus take our results in terms of temperature profiles and corresponding abundance profiles. We use the radiative transfer model already described in \citet{Cavalie2008b,Cavalie2013} to compute synthetic spectra of Uranus and Neptune in the mm-cm range. We extend our tropospheric thermal profiles of Uranus and Neptune from 2\,bar to 10\,mbar with the data from \citet{Orton2014a} and from the Herschel Science Centre ``ESA5'' model\footnote{http://archives.esac.esa.int/hsa/legacy/ADP/PlanetaryModels/} profiles for Uranus and Neptune, respectively. We add profiles of H$_2$S and NH$_3$ based on \citet{DeBoer1994} to account for their respective opacities, in addition to the microwave opacity of the H$_2$O line at 22\,GHz. The NH$_3$ mole fraction between the NH$_4$SH and NH$_3$ clouds is set to 6\dix{-6} and 1\dix{-5} in Uranus and Neptune (\citealt{Moreno1998} and ``ESA5'' model, resp.). Below the NH$_4$SH cloud, the NH$_3$ mole fraction is set to 1\dix{-4} and the deep H$_2$S mole fraction is set such that all H$_2$S is consumed to form the NH$_4$SH cloud. We use \citet{Bellotti2016} for the H$_2$O microwave opacity, \citet{Moreno1998} for the NH$_3$ microwave opacity, \citet{Liebe1969} for H$_2$- and He-broadening and temperature dependence parameters for the 22\,GHz line of H$_2$O, \citet{Fletcher2007} for the H$_2$-broadening and temperature dependence parameters of NH$_3$ lines, \citet{DeBoer1994} for the H$_2$S microwave opacity and broadening parameters, and \citet{dePater1991} and \citet{Fray2009} for H$_2$S condensation laws. Other spectral line parameters come from the JPL Molecular Spectroscopy Database \citep{Pickett1998}. The collision-induced absorption caused by H$_2$-H$_2$, H$_2$-He, and H$_2$-CH$_4$ pairs is computed following \citet{Borysow1985,Borysow1988} and \citet{Borysow1986}. 
  
\begin{figure}[htp]
  \begin{center}
  \includegraphics[width=10cm,keepaspectratio]{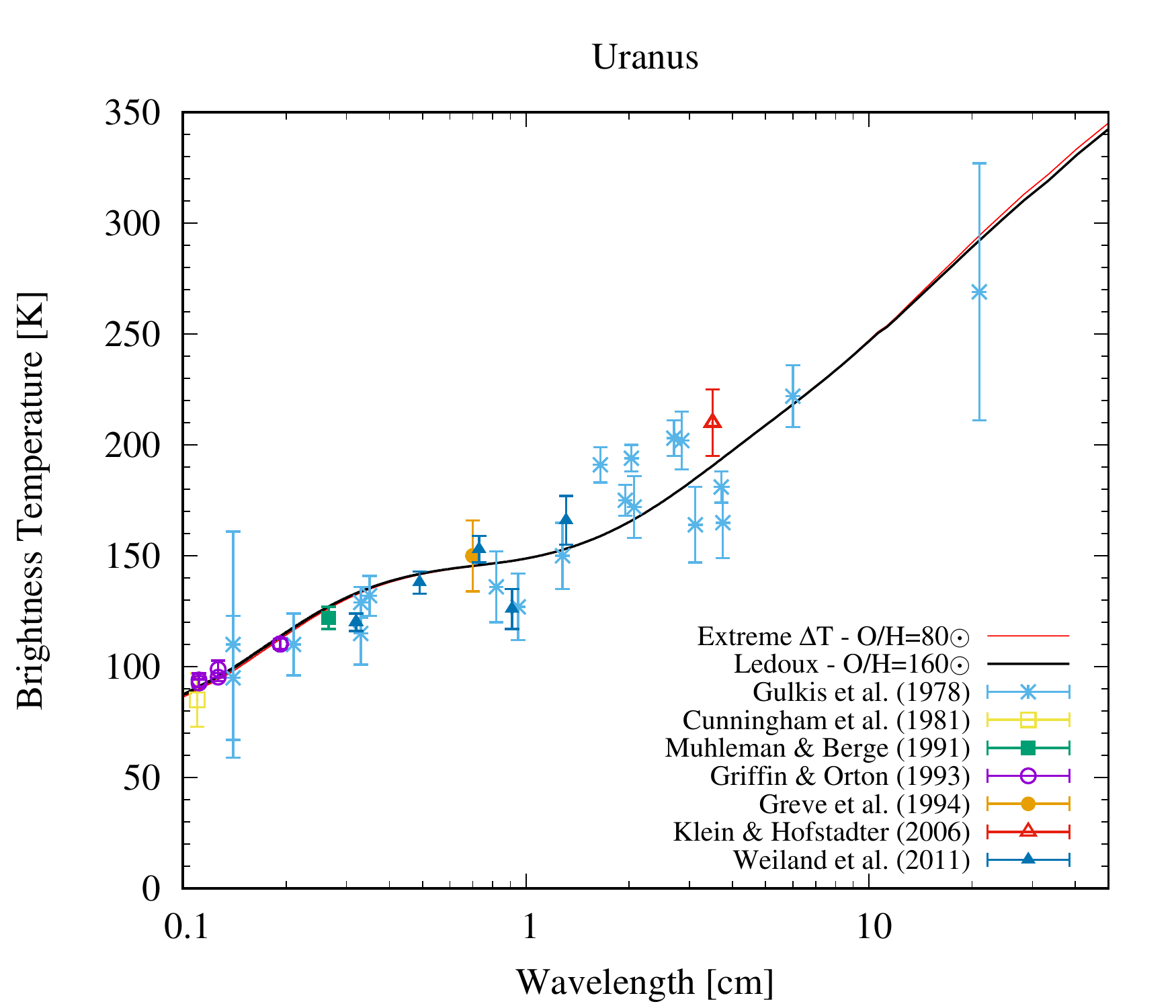}
  \includegraphics[width=10cm,keepaspectratio]{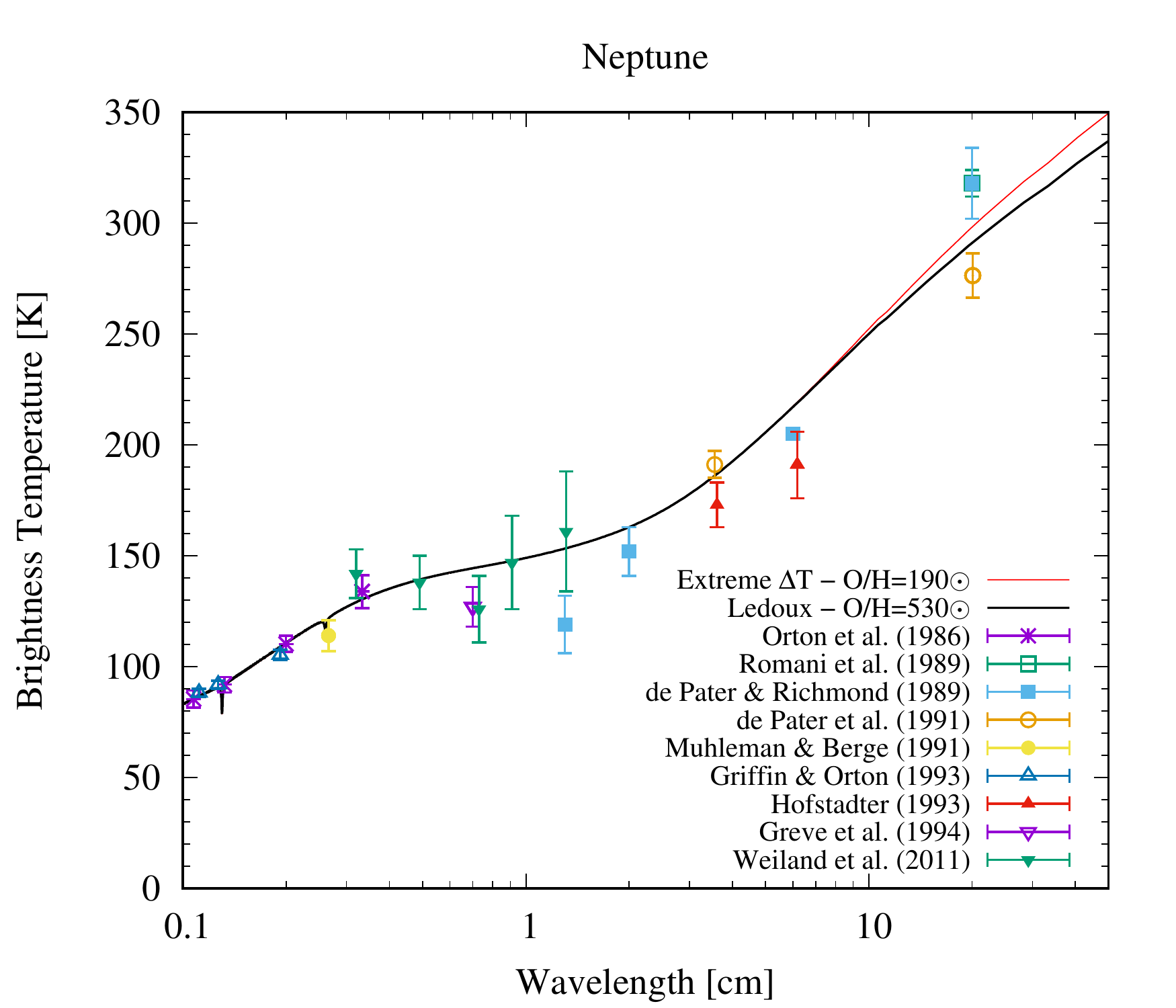}
  \caption{Brightness temperature spectra of Uranus and Neptune in the centimeter wavelength range. We account for the opacities of NH$_3$, H$_2$S, CO, and H$_2$O, on top of the collision-induced absorption spectrum of H$_2$-H$_2$, H$_2$-He, and H$_2$-CH$_4$. The two millimetric absorption lines in the Neptune spectrum are caused by CO. The spectral resolution is 1\,GHz. }
  \label{Ura_Nep_cm_spectra}
  \end{center}
\end{figure}

  The resulting spectra for Uranus and Neptune in the centimeter wavelength range, when considering the nominal cases of Section~\ref{Results}, are displayed in \fig{Ura_Nep_cm_spectra}. Even though we make no particular attempt to improve the fit to the data (especially around 1\,cm) by modifying the NH$_3$ and/or H$_2$S abundances, the obtained spectra are in good agreement with observations presented in \citet{Gulkis1978}, \citet{Cunningham1981}, \citet{Muhleman1991}, \citet{Griffin1993}, \citet{Greve1994}, \citet{Klein2006}, and \citet{Weiland2011}, for Uranus, and \citet{Orton1986}, \citet{Romani1989}, \citet{dePater1989}, \citet{dePater1991}, \citet{Muhleman1991}, \citet{Griffin1993}
, \citet{Hofstadter1993}, \citet{Greve1994}, and \citet{Weiland2011}, for Neptune. Even the extreme cases, in which the O abundance is lower but compensated by the biggest temperature jump allowed by the model (see Section~\ref{Temp_jump}), the cm-wave spectrum of both planets remains in agreement with observations. This comes from the fact that the region where the jump in temperature and H$_2$O abundance occur, i.e. the radiative layer, is only marginally probed by the H$_2$O absorption at 20\,cm wavelength, as shown in \citet{DeBoer1996}.

\section{Limitations and other questions \label{Limitations}}
In this section, we will detail the limitations of our model and stress the data that are desirable for increasing the predictability of such models.
    
    \subsection{The radiative layer as an insulation layer? \label{insulation_layer}}
    One of the novel aspects of our work is the inclusion of the stability criterion of \citet{Leconte2017} when extrapolating measured tropospheric temperatures to deeper levels where the atmosphere is in thermochemical equilibrium. Accounting for the mean molecular weight gradient when extending thermal profiles from observed levels in the upper troposphere to deeper levels implies the formation of a thin radiative layer ($\Delta z\sim$1\,km) in Uranus and Neptune at the levels where H$_2$O condenses. If such a stable layer exists, it should act as an insulation layer for the both the transport of heat and the chemical mixing, making our assumption of a vertically uniform $K_{zz}$ profile questionable. There is no estimate for chemical mixing under such conditions, but within this layer $K_{zz}$ could, in principle, be as low as that of molecular diffusivity $D$, resulting in a gradient for CO, a species that has a non-zero flux at the boundaries of the model. According to our model, molecular diffusivity is in the range of 10$^{-2}$-10$^{-3}$\Kunit, where the radiative layer in both planets resides. We estimate the variation in the CO mole fraction between the two limits of the radiative layer using \Flux{CO}$=$$-nD\derive{y_{\mathrm{CO}}}{t}$, where \Flux{CO} is the flux of CO at the upper boundary of the model (i.e., the top of the troposphere) and $n$ is the total number density. As CO is relatively stable in the stratospheres of Uranus and Neptune, a good approximation for \Flux{CO} is given by the input fluxes caused by external sources, and we take 10$^5$ and 10$^8$\Fluxunit~respectively \citep{Cavalie2014,Lellouch2005}. We confirm such values running test cases in the stratosphere with the photochemical model of \citet{Dobrijevic2010}. In this case only, the variation in the CO mole fraction between the two limits of the radiative layer would be of the order of magnitude of the observed $y_{\mathrm{CO}}^{\mathrm{top}}$. This means that the tropospheric CO could be partly caused by the external source because of the transport boundary caused by the radiative zone. Then, it would be impossible to constrain the deep oxygen abundance from the sole observation of $y_{\mathrm{CO}}^{\mathrm{top}}$ and thermochemical modeling, as presented in this paper. A full model accounting for thermochemistry in the troposphere, photochemistry in the stratosphere and external sources would be required to do so. This potentially is a significant limitation that should be kept in mind, although it only applies in the case where $K_{zz}$ is as low as $D$ in the radiative layer and the flux of CO from the stratosphere is as strong as the external source of CO. As long as $K_{zz}\geq10^{-1}$, the gradient disappears and $y_{\mathrm{CO}}^{\mathrm{top}}$ can be taken as a probe to the deep oxygen abundance. In 3D, overshooting across the 1\,km radiative layer is plausible and would help in this sense.

    \subsection{Chemical scheme\label{Chemical_scheme}}
    \citet{Moses2014} compared her chemical scheme with the one we have adopted in this work (from \citealt{Venot2012}) for studying the composition of HD189733b, a hot Jupiter. In her chemical scheme, she used the reaction rates updated by \citet{Moses2011}, based on \textit{ab initio} calculations. She found that the main difference between the two schemes resided in the reaction rate used for CH$_3$OH $+$ H $=$ CH$_3$ $+$ H$_2$O. \citet{Moses2011} then proceeded with a much slower reaction than \citet{Venot2012}. The difference between these two reaction rates, when used in the \citet{Venot2012} model has been shown for Jupiter and Saturn by \citet{Wang2016}. The scheme we use nominally in this study produces 10-30 times less CO than the same scheme with the \citet{Moses2011} rate for the CH$_3$OH $+$ H $=$ CH$_3$ $+$ H$_2$O reaction (see also their figures 4 and 5). 
    
    If we proceed as \citet{Wang2016} and alter in the \citet{Venot2012} scheme the reaction rate for CH$_3$OH $+$ H $=$ CH$_3$ $+$ H$_2$O by using the value of \citet{Moses2011}, and still use ``3-layer'' thermal profiles, we find that the (O/H,C/H) ratios in Uranus and Neptune become $\sim$ ($<$55,85) and $\sim$ (280,65) times solar, respectively. These values are significantly lower than with our nominal chemical scheme, as expected. However, we emphasize that these values are just presented for the sake of comparison. While the \textit{ab initio} calculations of \citet{Moses2011} are correct, including this rate in the scheme of \citet{Venot2012} as prescribed by \citet{Moses2014} in the chemical scheme does not enable reproducing of the experimental data of different combustion studies \citep{Cathonnet1982,Hidaka1989,Held1998} (R. Bounaceur, priv. comm. June 2016). Using such a reaction rate is thus not validated in the frame of our chemical scheme. In any case, it certainly underlines the urgent need for a better understanding of the kinetics of CH$_3$OH at high temperatures and pressures.

    \subsection{Tropospheric equation of state}
    In this work, we have implicitly used the ideal gas law. Significant water abundances at high pressures obviously render such a choice questionable. \citet{Nettelmann2008} proposed a new equation of state for H$_2$-He-H$_2$O mixtures, but for a higher pressure/temperature regime than investigated in this paper. More recently, \citet{Karpowicz2013} have proposed a tropospheric equation of state for Jupiter's troposphere. The effect on pressure and temperature seems rather limited (2\% decrease of T, compared to the ideal gas law, at their lower boundary of 200 bars), when compared to the temperature increases caused by the mean molecular weight gradient effect shown in this paper. Nevertheless, it remains to be seen what the effect of applying such equation of state to Uranus and Neptune would be, as they have presumably much more H$_2$O in their deep tropospheres than Jupiter. Unfortunately, the authors underline that their ``equation of state will likely have decreased accuracy under jovian conditions for pressures exceeding much beyond 100 bars, and may be invalid at pressures exceeding 2500 bars''. New laboratory experiments and theoretical simulations are thus needed to enable applying reliable equations of state for H$_2$-H$_2$O-He in the pressure/temperature range we are interested in (see \citealt{Baraffe2014} for a review). However, even if we had a robust non-ideal equation of state relevant for our conditions, our chemical model would need to be updated. Indeed, though we explicitly assume an ideal gas law in many calculations of our models, we also implicitly assume it in the kinetic rates.

  \subsection{Latent heat release by other condensates}
  In this work, we do not account for the release of latent heat caused by the formation of various clouds besides the H$_2$O clouds. Other expected clouds, from the deep troposphere to the upper troposphere, are supposedly composed of NH$_4$SH, NH$_3$ or H$_2$S (depending on which one is the most abundant in the deep troposphere and ``survives'' the formation of the NH$_4$SH cloud), and CH$_4$. Our choice to start our thermal profiles below the CH$_4$ cloud deck is meant so as not to have to account for the CH$_4$ cloud formation, related latent heat release, and even possible mean molecular weight gradient effect \citep{Guillot1995}.

\section{Conclusion \label{Conclusion}}
In this paper, we have modeled the thermochemistry in the tropospheres of Uranus and Neptune, in an attempt to constrain their deep oxygen abundance from upper tropospheric observations of CO. The derivation of the deep O/H ratio in Giant Planet tropospheres from thermochemical computations requires precise measurements of upper tropospheric CO and CH$_4$, knowledge of oxygen species kinetics, as well as a good knowledge of tropospheric temperatures and vertical mixing.

We have shown that the transition between a water-rich and a water-poor region in giant planet tropospheres results in a significant gradient of mean molecular weight that impacts, in turn, the shape of the thermal profile. Accounting for this gradient in the condensation zone of H$_2$O by applying the stability criterion of \citet{Leconte2017} implies a significant departure of temperature from the dry/wet adiabats that have been used so far in previous papers. Our results show that large oxygen enrichments can produce a thin radiative layer ($\Delta z\sim$1\,km) where H$_2$O condenses. The thermal gradient, stabilized by the downward increasing mean molecular weight, gets very strong and becomes radiative in this zone. This results in higher temperatures in deep levels compared to dry/wet adiabats. 

While the mean molecular weight gradient effects are beginning to be applied to internal and formation models of giant planets \citep{Leconte2012,Vazan2015}, our model shows for the first time the importance of accounting for these effects already at upper tropospheric levels for oxygen-rich giant planets \citep{Leconte2017}. Our results show that these new profiles lower the required oxygen abundances to reproduce the CO observations. However, our estimates are affected by the current limitations of the model, like the lack of Rosseland opacities for the considered mixtures. Another source of uncertainty is related to the high pressures at the quench level which would require non ideal equation of states and modified kinetics. While a better equation of state can be expected soon and is rather simple to implement, its impact on the kinetics might be extremely difficult to derive. Finally, it is noticeable that if (but only if) vertical mixing is as low as molecular diffusivity in the $\sim$1\,km thin radiative layer, then it is much more complicated to constrain the deep oxygen abundance in Uranus and Neptune because the CO external source then needs to be precisely characterized as well.

While using such temperature models should lead to lower O/H ratios than when using dry/wet adiabats, our nominal value for Neptune's O/H ratio is close to the value derived by \citet{Luszcz-Cook2013} (they have used wet adiabats). This is caused by differences in oxygen chemistry. This certainly underlines the need for improved knowledge of the oxygen species thermochemistry. For instance, with kinetics that reduce significantly the destruction of CH$_3$OH, as proposed by \citet{Moses2011} and applied by \citet{Wang2016}, then the O/H required to fit the upper tropospheric observations of CO significantly decreases. However, using such different kinetics first require validation within a complete scheme over the temperature/pressure range relevant to such studies. We believe this is the main strength of the chemical scheme we are using, but we do not underestimate the work that still needs to be done to improve our knowledge of CH$_3$OH kinetics.

In our nominal model, in which we fix the temperature, the CH$_4$ abundance, and the tropospheric mixing to our best estimates, the O/H in Uranus and Neptune are $<$160 and 540 times the solar value, respectively. While Uranus' and Neptune's ices formation and heavy element enrichment could result from clathration, this scenario would imply different mixtures of CO/CO$_2$ at the time of planetesimal condensation. The difference seen in their C/O ratios ($>$0.23 for Uranus and $\sim$0.03 for Neptune) is actually rather puzzling and may therefore be an indication that convection is strongly inhibited in the troposphere of Uranus. Indeed, this appealing explanation could enable reconciling formation models for the two planets with observations of their tropospheric composition. Among other things, the inconsistency between Uranus' and Neptune's O/H remains puzzling for now, and new data from dedicated orbiters with atmospheric descent probes are highly desirable to study these mysterious worlds \citep{Arridge2012,Arridge2014,Masters2014,Turrini2014}. In the meantime, \textit{James Webb Space Telescope} observations will enable measurements that will better constrain the deep abundances of several heavy elements, like Ge, As, and P \citep{Norwood2016a,Norwood2016b}.

Strong uncertainties remain in Uranus and Neptune on the upper tropospheric CH$_4$ mole fraction, on the tropospheric mixing and on the tropospheric temperature profile. Because these are central to such thermochemical modeling, we have explored this whole parameter space with a chemical model that uses a scheme validated in the temperature/pressure range of interest here. When more precise measurements of temperature, mixing and CH$_4$ abundance are obtained, our results can be used to find the corresponding O/H required to fit the tropospheric CO in both planets.

In the case of the gas giants, the lower oxygen enrichments that are expected should result in a much less significant mean molecular weight gradient in the condensation region of H$_2$O and thus a less spectacular effect on the thermal profile and on the derivation of their deep O/H ratio. In this sense, it will be interesting to compare our thermochemical model results with the measurements that will be provided by Juno \citep{Matousek2007,Helled2014}, which successfully performed its Jupiter orbit insertion on July 4, 2016. At Saturn, a descent probe has recently been proposed for an ESA M5 mission \citep{Mousis2014,Mousis2016}. While this probe may not be able to reach the well-mixed region of H$_2$O and thus measure Saturn's O/H, it would provide ground truth as to the abundances of other heavy elements and disequilibrium species and shed light on the formation processes of Saturn.


\section*{Acknowledgements}
  T. Cavali\'e thanks the CNRS/INSU Programme National de Plan\'etologie (PNP) for funding support. O. Venot acknowledges support from the KU Leuven IDO project IDO/10/2013, from the FWO Postdoctoral Fellowship program, and from the Centre National d'\'Etudes Spatiales (CNES). T. Cavali\'e and O.Venot thank R. Bounaceur (LRPG, Nancy, France) for useful discussions on the kinetics of CH$_3$OH. T. Cavali\'e thanks I. de Pater (Univ. of California, Berkeley, USA) for providing her H$_2$S condensation laws. We thank Dr. Dong Wang and an anonymous reviewer for they constructive comments that help us improve our paper.


\section*{References}
\bibliographystyle{elsarticle-harv} 

\begin{thebibliography}{144}
\expandafter\ifx\csname natexlab\endcsname\relax\def\natexlab#1{#1}\fi
\expandafter\ifx\csname url\endcsname\relax
  \def\url#1{\texttt{#1}}\fi
\expandafter\ifx\csname urlprefix\endcsname\relax\def\urlprefix{URL }\fi

\bibitem[{{Ag{\'u}ndez} et~al.(2014){Ag{\'u}ndez}, {Parmentier}, {Venot},
  {Hersant}, and {Selsis}}]{Agundez2014}
{Ag{\'u}ndez}, M., {Parmentier}, V., {Venot}, O., {Hersant}, F., {Selsis}, F.,
  2014. {Pseudo 2D chemical model of hot-Jupiter atmospheres: application to HD
  209458b and HD 189733b}. \aap 564, A73.

\bibitem[{{Ali-Dib} et~al.(2014){Ali-Dib}, {Mousis}, {Petit}, and
  {Lunine}}]{Ali-Dib2014}
{Ali-Dib}, M., {Mousis}, O., {Petit}, J.-M., {Lunine}, J.~I., 2014. {The
  Measured Compositions of Uranus and Neptune from their Formation on the CO
  Ice Line}. \apj 793, 9.

\bibitem[{{Anders} and {Grevesse}(1989)}]{Anders1989}
{Anders}, E., {Grevesse}, N., 1989. {Abundances of the elements - Meteoritic
  and solar}. \gca 53, 197--214.

\bibitem[{{Arridge} et~al.(2012){Arridge}, {Agnor}, {Andr{\'e}}, {Baines},
  {Fletcher}, {Gautier}, {Hofstadter}, {Jones}, {Lamy}, {Langevin}, {Mousis},
  {Nettelmann}, {Russell}, {Stallard}, {Tiscareno}, {Tobie}, {Bacon},
  {Chaloner}, {Guest}, {Kemble}, {Peacocke}, {Achilleos}, {Andert}, {Banfield},
  {Barabash}, {Barthelemy}, {Bertucci}, {Brandt}, {Cecconi}, {Chakrabarti},
  {Cheng}, {Christensen}, {Christou}, {Coates}, {Collinson}, {Cooper},
  {Courtin}, {Dougherty}, {Ebert}, {Entradas}, {Fazakerley}, {Fortney},
  {Galand}, {Gustin}, {Hedman}, {Helled}, {Henri}, {Hess}, {Holme},
  {Karatekin}, {Krupp}, {Leisner}, {Martin-Torres}, {Masters}, {Melin},
  {Miller}, {M{\"u}ller-Wodarg}, {Noyelles}, {Paranicas}, {de Pater},
  {P{\"a}tzold}, {Prang{\'e}}, {Qu{\'e}merais}, {Roussos}, {Rymer},
  {S{\'a}nchez-Lavega}, {Saur}, {Sayanagi}, {Schenk}, {Schubert}, {Sergis},
  {Sohl}, {Sittler}, {Teanby}, {Tellmann}, {Turtle}, {Vinatier}, {Wahlund}, and
  {Zarka}}]{Arridge2012}
{Arridge}, C.~S., {Agnor}, C.~B., {Andr{\'e}}, N., {Baines}, K.~H., {Fletcher},
  L.~N., {Gautier}, D., {Hofstadter}, M.~D., {Jones}, G.~H., {Lamy}, L.,
  {Langevin}, Y., {Mousis}, O., {Nettelmann}, N., {Russell}, C.~T., {Stallard},
  T., {Tiscareno}, M.~S., {Tobie}, G., {Bacon}, A., {Chaloner}, C., {Guest},
  M., {Kemble}, S., {Peacocke}, L., {Achilleos}, N., {Andert}, T.~P.,
  {Banfield}, D., {Barabash}, S., {Barthelemy}, M., {Bertucci}, C., {Brandt},
  P., {Cecconi}, B., {Chakrabarti}, S., {Cheng}, A.~F., {Christensen}, U.,
  {Christou}, A., {Coates}, A.~J., {Collinson}, G., {Cooper}, J.~F., {Courtin},
  R., {Dougherty}, M.~K., {Ebert}, R.~W., {Entradas}, M., {Fazakerley}, A.~N.,
  {Fortney}, J.~J., {Galand}, M., {Gustin}, J., {Hedman}, M., {Helled}, R.,
  {Henri}, P., {Hess}, S., {Holme}, R., {Karatekin}, {\"O}., {Krupp}, N.,
  {Leisner}, J., {Martin-Torres}, J., {Masters}, A., {Melin}, H., {Miller}, S.,
  {M{\"u}ller-Wodarg}, I., {Noyelles}, B., {Paranicas}, C., {de Pater}, I.,
  {P{\"a}tzold}, M., {Prang{\'e}}, R., {Qu{\'e}merais}, E., {Roussos}, E.,
  {Rymer}, A.~M., {S{\'a}nchez-Lavega}, A., {Saur}, J., {Sayanagi}, K.~M.,
  {Schenk}, P., {Schubert}, G., {Sergis}, N., {Sohl}, F., {Sittler}, E.~C.,
  {Teanby}, N.~A., {Tellmann}, S., {Turtle}, E.~P., {Vinatier}, S., {Wahlund},
  J.-E., {Zarka}, P., 2012. {Uranus Pathfinder: exploring the origins and
  evolution of Ice Giant planets}. \expastr 33, 753--791.

\bibitem[{{Arridge} et~al.(2014){Arridge}, {Achilleos}, {Agarwal}, {Agnor},
  {Ambrosi}, {Andr{\'e}}, {Badman}, {Baines}, {Banfield}, {Barth{\'e}l{\'e}my},
  {Bisi}, {Blum}, {Bocanegra-Bahamon}, {Bonfond}, {Bracken}, {Brandt},
  {Briand}, {Briois}, {Brooks}, {Castillo-Rogez}, {Cavali{\'e}}, {Christophe},
  {Coates}, {Collinson}, {Cooper}, {Costa-Sitja}, {Courtin}, {Daglis}, {de
  Pater}, {Desai}, {Dirkx}, {Dougherty}, {Ebert}, {Filacchione}, {Fletcher},
  {Fortney}, {Gerth}, {Grassi}, {Grodent}, {Gr{\"u}n}, {Gustin}, {Hedman},
  {Helled}, {Henri}, {Hess}, {Hillier}, {Hofstadter}, {Holme}, {Horanyi},
  {Hospodarsky}, {Hsu}, {Irwin}, {Jackman}, {Karatekin}, {Kempf}, {Khalisi},
  {Konstantinidis}, {Kr{\"u}ger}, {Kurth}, {Labrianidis}, {Lainey}, {Lamy},
  {Laneuville}, {Lucchesi}, {Luntzer}, {MacArthur}, {Maier}, {Masters},
  {McKenna-Lawlor}, {Melin}, {Milillo}, {Moragas-Klostermeyer}, {Morschhauser},
  {Moses}, {Mousis}, {Nettelmann}, {Neubauer}, {Nordheim}, {Noyelles}, {Orton},
  {Owens}, {Peron}, {Plainaki}, {Postberg}, {Rambaux}, {Retherford}, {Reynaud},
  {Roussos}, {Russell}, {Rymer}, {Sallantin}, {S{\'a}nchez-Lavega}, {Santolik},
  {Saur}, {Sayanagi}, {Schenk}, {Schubert}, {Sergis}, {Sittler}, {Smith},
  {Spahn}, {Srama}, {Stallard}, {Sterken}, {Sternovsky}, {Tiscareno}, {Tobie},
  {Tosi}, {Trieloff}, {Turrini}, {Turtle}, {Vinatier}, {Wilson}, and
  {Zarka}}]{Arridge2014}
{Arridge}, C.~S., {Achilleos}, N., {Agarwal}, J., {Agnor}, C.~B., {Ambrosi},
  R., {Andr{\'e}}, N., {Badman}, S.~V., {Baines}, K., {Banfield}, D.,
  {Barth{\'e}l{\'e}my}, M., {Bisi}, M.~M., {Blum}, J., {Bocanegra-Bahamon}, T.,
  {Bonfond}, B., {Bracken}, C., {Brandt}, P., {Briand}, C., {Briois}, C.,
  {Brooks}, S., {Castillo-Rogez}, J., {Cavali{\'e}}, T., {Christophe}, B.,
  {Coates}, A.~J., {Collinson}, G., {Cooper}, J.~F., {Costa-Sitja}, M.,
  {Courtin}, R., {Daglis}, I.~A., {de Pater}, I., {Desai}, M., {Dirkx}, D.,
  {Dougherty}, M.~K., {Ebert}, R.~W., {Filacchione}, G., {Fletcher}, L.~N.,
  {Fortney}, J., {Gerth}, I., {Grassi}, D., {Grodent}, D., {Gr{\"u}n}, E.,
  {Gustin}, J., {Hedman}, M., {Helled}, R., {Henri}, P., {Hess}, S., {Hillier},
  J.~K., {Hofstadter}, M.~H., {Holme}, R., {Horanyi}, M., {Hospodarsky}, G.,
  {Hsu}, S., {Irwin}, P., {Jackman}, C.~M., {Karatekin}, O., {Kempf}, S.,
  {Khalisi}, E., {Konstantinidis}, K., {Kr{\"u}ger}, H., {Kurth}, W.~S.,
  {Labrianidis}, C., {Lainey}, V., {Lamy}, L.~L., {Laneuville}, M., {Lucchesi},
  D., {Luntzer}, A., {MacArthur}, J., {Maier}, A., {Masters}, A.,
  {McKenna-Lawlor}, S., {Melin}, H., {Milillo}, A., {Moragas-Klostermeyer}, G.,
  {Morschhauser}, A., {Moses}, J.~I., {Mousis}, O., {Nettelmann}, N.,
  {Neubauer}, F.~M., {Nordheim}, T., {Noyelles}, B., {Orton}, G.~S., {Owens},
  M., {Peron}, R., {Plainaki}, C., {Postberg}, F., {Rambaux}, N., {Retherford},
  K., {Reynaud}, S., {Roussos}, E., {Russell}, C.~T., {Rymer}, A.~M.,
  {Sallantin}, R., {S{\'a}nchez-Lavega}, A., {Santolik}, O., {Saur}, J.,
  {Sayanagi}, K.~M., {Schenk}, P., {Schubert}, J., {Sergis}, N., {Sittler},
  E.~C., {Smith}, A., {Spahn}, F., {Srama}, R., {Stallard}, T., {Sterken}, V.,
  {Sternovsky}, Z., {Tiscareno}, M., {Tobie}, G., {Tosi}, F., {Trieloff}, M.,
  {Turrini}, D., {Turtle}, E.~P., {Vinatier}, S., {Wilson}, R., {Zarka}, P.,
  2014. {The science case for an orbital mission to Uranus: Exploring the
  origins and evolution of ice giant planets}. \planss 104, 122--140.

\bibitem[{{Atkinson} et~al.(2016){Atkinson}, {Simon}, {Banfield}, {Atreya},
  {Blacksberg}, {Brinckerhoff}, {Colaprete}, {Coustenis}, {Fletcher},
  {Guillot}, {Hofstadter}, {Lunine}, {Mahaffy}, {Marley}, {Mousis}, {Spilker},
  {Trainer}, and {Webster}}]{Atkinson2016}
{Atkinson}, D.~H., {Simon}, A.~A., {Banfield}, D., {Atreya}, S.~K.,
  {Blacksberg}, J., {Brinckerhoff}, W., {Colaprete}, A., {Coustenis}, A.,
  {Fletcher}, L., {Guillot}, T., {Hofstadter}, M., {Lunine}, J.~I., {Mahaffy},
  P., {Marley}, M.~S., {Mousis}, O., {Spilker}, T.~R., {Trainer}, M.~G.,
  {Webster}, C., 2016. {Exploring Saturn - The Saturn PRobe Interior and
  aTmosphere Explorer (SPRITE) Mission}. In: AAS/Division for Planetary
  Sciences Meeting Abstracts. Vol.~48 of AAS/Division for Planetary Sciences
  Meeting Abstracts. p. \#123.29.

\bibitem[{{Atreya} et~al.(1999){Atreya}, {Wong}, {Owen}, {Mahaffy}, {Niemann},
  {de Pater}, {Drossart}, and {Encrenaz}}]{Atreya1999}
{Atreya}, S.~K., {Wong}, M.~H., {Owen}, T.~C., {Mahaffy}, P.~R., {Niemann},
  H.~B., {de Pater}, I., {Drossart}, P., {Encrenaz}, T., 1999. {A comparison of
  the atmospheres of Jupiter and Saturn: deep atmospheric composition, cloud
  structure, vertical mixing, and origin}. \planss 47, 1243--1262.

\bibitem[{{Baines} et~al.(1995){Baines}, {Mickelson}, {Larson}, and
  {Ferguson}}]{Baines1995}
{Baines}, K.~H., {Mickelson}, M.~E., {Larson}, L.~E., {Ferguson}, D.~W., 1995.
  {The abundances of methane and ortho/para hydrogen on Uranus and Neptune:
  Implications of New Laboratory 4-0 H2 quadrupole line parameters}. \icarus
  114, 328--340.

\bibitem[{{Bar-Nun} et~al.(1988){Bar-Nun}, {Kleinfeld}, and
  {Kochavi}}]{Bar-Nun1988}
{Bar-Nun}, A., {Kleinfeld}, I., {Kochavi}, E., 1988. {Trapping of gas mixtures
  by amorphous water ice}. \prb 38, 7749--7754.

\bibitem[{{Baraffe} et~al.(2014){Baraffe}, {Chabrier}, {Fortney}, and
  {Sotin}}]{Baraffe2014}
{Baraffe}, I., {Chabrier}, G., {Fortney}, J., {Sotin}, C., 2014. {Planetary
  internal structures}. ArXiv e-prints.

\bibitem[{{Batalha} et~al.(2013){Batalha}, {Rowe}, {Bryson}, {Barclay},
  {Burke}, {Caldwell}, {Christiansen}, {Mullally}, {Thompson}, {Brown},
  {Dupree}, {Fabrycky}, {Ford}, {Fortney}, {Gilliland}, {Isaacson}, {Latham},
  {Marcy}, {Quinn}, {Ragozzine}, {Shporer}, {Borucki}, {Ciardi}, {Gautier},
  {Haas}, {Jenkins}, {Koch}, {Lissauer}, {Rapin}, {Basri}, {Boss}, {Buchhave},
  {Carter}, {Charbonneau}, {Christensen-Dalsgaard}, {Clarke}, {Cochran},
  {Demory}, {Desert}, {Devore}, {Doyle}, {Esquerdo}, {Everett}, {Fressin},
  {Geary}, {Girouard}, {Gould}, {Hall}, {Holman}, {Howard}, {Howell},
  {Ibrahim}, {Kinemuchi}, {Kjeldsen}, {Klaus}, {Li}, {Lucas}, {Meibom},
  {Morris}, {Pr{\v s}a}, {Quintana}, {Sanderfer}, {Sasselov}, {Seader},
  {Smith}, {Steffen}, {Still}, {Stumpe}, {Tarter}, {Tenenbaum}, {Torres},
  {Twicken}, {Uddin}, {Van Cleve}, {Walkowicz}, and {Welsh}}]{Batalha2013}
{Batalha}, N.~M., {Rowe}, J.~F., {Bryson}, S.~T., {Barclay}, T., {Burke},
  C.~J., {Caldwell}, D.~A., {Christiansen}, J.~L., {Mullally}, F., {Thompson},
  S.~E., {Brown}, T.~M., {Dupree}, A.~K., {Fabrycky}, D.~C., {Ford}, E.~B.,
  {Fortney}, J.~J., {Gilliland}, R.~L., {Isaacson}, H., {Latham}, D.~W.,
  {Marcy}, G.~W., {Quinn}, S.~N., {Ragozzine}, D., {Shporer}, A., {Borucki},
  W.~J., {Ciardi}, D.~R., {Gautier}, III, T.~N., {Haas}, M.~R., {Jenkins},
  J.~M., {Koch}, D.~G., {Lissauer}, J.~J., {Rapin}, W., {Basri}, G.~S., {Boss},
  A.~P., {Buchhave}, L.~A., {Carter}, J.~A., {Charbonneau}, D.,
  {Christensen-Dalsgaard}, J., {Clarke}, B.~D., {Cochran}, W.~D., {Demory},
  B.-O., {Desert}, J.-M., {Devore}, E., {Doyle}, L.~R., {Esquerdo}, G.~A.,
  {Everett}, M., {Fressin}, F., {Geary}, J.~C., {Girouard}, F.~R., {Gould}, A.,
  {Hall}, J.~R., {Holman}, M.~J., {Howard}, A.~W., {Howell}, S.~B., {Ibrahim},
  K.~A., {Kinemuchi}, K., {Kjeldsen}, H., {Klaus}, T.~C., {Li}, J., {Lucas},
  P.~W., {Meibom}, S., {Morris}, R.~L., {Pr{\v s}a}, A., {Quintana}, E.,
  {Sanderfer}, D.~T., {Sasselov}, D., {Seader}, S.~E., {Smith}, J.~C.,
  {Steffen}, J.~H., {Still}, M., {Stumpe}, M.~C., {Tarter}, J.~C., {Tenenbaum},
  P., {Torres}, G., {Twicken}, J.~D., {Uddin}, K., {Van Cleve}, J.,
  {Walkowicz}, L., {Welsh}, W.~F., 2013. {Planetary Candidates Observed by
  Kepler. III. Analysis of the First 16 Months of Data}. \apjs 204, 24.

\bibitem[{{Beer}(1975)}]{Beer1975}
{Beer}, R., 1975. {Detection of carbon monoxide in Jupiter}. \apjl 200,
  L167--L169.

\bibitem[{{Bellotti} et~al.(2016){Bellotti}, {Steffes}, and
  {Chinsomboom}}]{Bellotti2016}
{Bellotti}, A., {Steffes}, P.~G., {Chinsomboom}, G., 2016. {Laboratory
  measurements of the 5-20 cm wavelength opacity of ammonia, water vapor, and
  methane under simulated conditions for the deep jovian atmosphere}. \icarus
  280, 255--267.

\bibitem[{{B{\'e}zard} et~al.(2002){B{\'e}zard}, {Lellouch}, {Strobel},
  {Maillard}, and {Drossart}}]{Bezard2002}
{B{\'e}zard}, B., {Lellouch}, E., {Strobel}, D., {Maillard}, J.-P., {Drossart},
  P., 2002. {Carbon Monoxide on Jupiter: Evidence for Both Internal and
  External Sources}. \icarus 159, 95--111.

\bibitem[{{Borysow} et~al.(1985){Borysow}, {Trafton}, {Frommhold}, and
  {Birnbaum}}]{Borysow1985}
{Borysow}, J., {Trafton}, L., {Frommhold}, L., {Birnbaum}, G., 1985. {Modeling
  of pressure-induced far-infrared absorption spectra Molecular hydrogen
  pairs}. \apj 296, 644--654.

\bibitem[{{Borysow} and {Frommhold}(1986)}]{Borysow1986}
{Borysow}, A., {Frommhold}, L., 1986. {Theoretical collision-induced
  rototranslational absorption spectra for the outer planets - H$_2$-CH$_4$
  pairs}. \apj 304, 849--865.

\bibitem[{{Borysow} et~al.(1988){Borysow}, {Frommhold}, and
  {Birnbaum}}]{Borysow1988}
{Borysow}, J., {Frommhold}, L., {Birnbaum}, G., 1988. {Collison-induced
  rototranslational absorption spectra of H$_2$-He pairs at temperatures from
  40 to 3000 K}. \apj 326, 509--515.

\bibitem[{{Boss}(1997)}]{Boss1997}
{Boss}, A.~P., 1997. {Giant planet formation by gravitational instability.}
  Science 276, 1836--1839.

\bibitem[{{Bounaceur} et~al.(2007){Bounaceur}, {Glaude}, {Fournet},
  {Battin-Leclerc}, {Jay}, and {Pires Da Cruz}}]{Bounaceur2007}
{Bounaceur}, R., {Glaude}, P.~A., {Fournet}, R., {Battin-Leclerc}, F., {Jay},
  S., {Pires Da Cruz}, A., 2007. {Kinetic modelling of a surrogate diesel fuel
  applied to 3D auto-ignition in HCCI engines.} Int. J. Vehicle Design 44, 124
  -- 142.

\bibitem[{{Burgdorf} et~al.(2003){Burgdorf}, {Orton}, {Davis}, {Sidher},
  {Feuchtgruber}, {Griffin}, and {Swinyard}}]{Burgdorf2003}
{Burgdorf}, M., {Orton}, G.~S., {Davis}, G.~R., {Sidher}, S.~D.,
  {Feuchtgruber}, H., {Griffin}, M.~J., {Swinyard}, B.~M., 2003. {Neptune's
  far-infrared spectrum from the ISO long-wavelength and short-wavelength
  spectrometers}. \icarus 164, 244--253.

\bibitem[{{Cathonnet} et~al.(1982){Cathonnet}, {Boettner}, and
  {James}}]{Cathonnet1982}
{Cathonnet}, M., {Boettner}, J.~C., {James}, H., 1982. {Study of methanol
  oxidation and self ignition in the temperature-range 500-600-degrees-C}.
  Journal de Chimie Physique Et De Physico-Chimie Biologique 79, 475--478.

\bibitem[{{Cavali{\'e}} et~al.(2008{\natexlab{a}}){Cavali{\'e}}, {Billebaud},
  {Biver}, {Dobrijevic}, {Lellouch}, {Brillet}, {Lecacheux}, {Hjalmarson},
  {Sandqvist}, {Frisk}, {Olberg}, {Bergin}, and {The Odin Team}}]{Cavalie2008b}
{Cavali{\'e}}, T., {Billebaud}, F., {Biver}, N., {Dobrijevic}, M., {Lellouch},
  E., {Brillet}, J., {Lecacheux}, A., {Hjalmarson}, {\AA}., {Sandqvist}, A.,
  {Frisk}, U., {Olberg}, M., {Bergin}, E.~A., {The Odin Team},
  2008{\natexlab{a}}. {Observation of water vapor in the stratosphere of
  Jupiter with the Odin space telescope}. \planss 56, 1573--1584.

\bibitem[{{Cavali{\'e}} et~al.(2008{\natexlab{b}}){Cavali{\'e}}, {Billebaud},
  {Fouchet}, {Lellouch}, {Brillet}, and {Dobrijevic}}]{Cavalie2008a}
{Cavali{\'e}}, T., {Billebaud}, F., {Fouchet}, T., {Lellouch}, E., {Brillet},
  J., {Dobrijevic}, M., 2008{\natexlab{b}}. {Observations of CO on Saturn and
  Uranus at millimeter wavelengths: new upper limit determinations}. \aap 484,
  555--561.

\bibitem[{{Cavali{\'e}} et~al.(2009){Cavali{\'e}}, {Billebaud}, {Dobrijevic},
  {Fouchet}, {Lellouch}, {Encrenaz}, {Brillet}, {Moriarty-Schieven},
  {Wouterloot}, and {Hartogh}}]{Cavalie2009}
{Cavali{\'e}}, T., {Billebaud}, F., {Dobrijevic}, M., {Fouchet}, T.,
  {Lellouch}, E., {Encrenaz}, T., {Brillet}, J., {Moriarty-Schieven}, G.~H.,
  {Wouterloot}, J.~G.~A., {Hartogh}, P., 2009. {First observation of CO at 345
  GHz in the atmosphere of Saturn with the JCMT: New constraints on its
  origin}. \icarus 203, 531--540.

\bibitem[{{Cavali{\'e}} et~al.(2013){Cavali{\'e}}, {Feuchtgruber}, {Lellouch},
  {de Val-Borro}, {Jarchow}, {Moreno}, {Hartogh}, {Orton}, {Greathouse},
  {Billebaud}, {Dobrijevic}, {Lara}, {Gonz{\'a}lez}, and
  {Sagawa}}]{Cavalie2013}
{Cavali{\'e}}, T., {Feuchtgruber}, H., {Lellouch}, E., {de Val-Borro}, M.,
  {Jarchow}, C., {Moreno}, R., {Hartogh}, P., {Orton}, G., {Greathouse}, T.~K.,
  {Billebaud}, F., {Dobrijevic}, M., {Lara}, L.~M., {Gonz{\'a}lez}, A.,
  {Sagawa}, H., 2013. {Spatial distribution of water in the stratosphere of
  Jupiter from Herschel HIFI and PACS observations}. \aap 553, A21.

\bibitem[{{Cavali{\'e}} et~al.(2014){Cavali{\'e}}, {Moreno}, {Lellouch},
  {Hartogh}, {Venot}, {Orton}, {Jarchow}, {Encrenaz}, {Selsis}, {Hersant}, and
  {Fletcher}}]{Cavalie2014}
{Cavali{\'e}}, T., {Moreno}, R., {Lellouch}, E., {Hartogh}, P., {Venot}, O.,
  {Orton}, G.~S., {Jarchow}, C., {Encrenaz}, T., {Selsis}, F., {Hersant}, F.,
  {Fletcher}, L.~N., 2014. {The first submillimeter observation of CO in the
  stratosphere of Uranus}. \aap 562, A33.

\bibitem[{{Conrath} et~al.(1987){Conrath}, {Hanel}, {Gautier}, {Marten}, and
  {Lindal}}]{Conrath1987}
{Conrath}, B., {Hanel}, R., {Gautier}, D., {Marten}, A., {Lindal}, G., 1987.
  {The helium abundance of Uranus from Voyager measurements}. \jgr 92,
  15003--15010.

\bibitem[{{Conrath} and {Gautier}(2000)}]{Conrath2000}
{Conrath}, B.~J., {Gautier}, D., 2000. {Saturn Helium Abundance: A Reanalysis
  of Voyager Measurements}. \icarus 144, 124--134.

\bibitem[{{Cunningham} et~al.(1981){Cunningham}, {Ade}, {Robson}, {Nolt}, and
  {Radostitz}}]{Cunningham1981}
{Cunningham}, C.~T., {Ade}, P.~A.~R., {Robson}, E.~I., {Nolt}, I.~G.,
  {Radostitz}, J.~V., 1981. {The submillimeter spectra of the planets -
  Narrow-band photometry}. \icarus 48, 127--139.

\bibitem[{{de Pater} and {Richmond}(1989)}]{dePater1989}
{de Pater}, I., {Richmond}, M., 1989. {Neptune's microwave spectrum from 1
  MM to 20 CM}. \icarus 80, 1--13.

\bibitem[{{de Pater} et~al.(1991){de Pater}, {Romani}, and
  {Atreya}}]{dePater1991}
{de Pater}, I., {Romani}, P.~N., {Atreya}, S.~K., 1991. {Possible microwave
  absorption by H2S gas in Uranus' and Neptune's atmospheres}. \icarus 91,
  220--233.

\bibitem[{{de Pater} et~al.(2005){de Pater}, {DeBoer}, {Marley}, {Freedman},
  and {Young}}]{dePater2005}
{de Pater}, I., {DeBoer}, D., {Marley}, M., {Freedman}, R., {Young}, R., 2005.
  {Retrieval of water in Jupiter's deep atmosphere using microwave spectra of
  its brightness temperature}. \icarus 173, 425--438.

\bibitem[{{DeBoer} and {Steffes}(1994)}]{DeBoer1994}
{DeBoer}, D.~R., {Steffes}, P.~G., 1994. {Laboratory measurements of the
  microwave properties of H2S under simulated Jovian conditions with an
  application to Neptune}. \icarus 109, 352--366.

\bibitem[{{DeBoer} and {Steffes}(1996)}]{DeBoer1996}
{DeBoer}, D.~R., {Steffes}, P.~G., 1996. {Estimates of the Tropospheric 
  Vertical Structure of Neptune Based on Microwave Radiative Transfer 
  Studies}. \icarus 123, 324--335.

\bibitem[{{Dobrijevic} et~al.(2010){Dobrijevic}, {Cavali{\'e}}, {H{\'e}brard},
  {Billebaud}, {Hersant}, and {Selsis}}]{Dobrijevic2010}
{Dobrijevic}, M., {Cavali{\'e}}, T., {H{\'e}brard}, E., {Billebaud}, F.,
  {Hersant}, F., {Selsis}, F., 2010. {Key reactions in the photochemistry
  of hydrocarbons in Neptune's stratosphere}. \planss 58, 1555--1566.

\bibitem[{{Fegley} and {Prinn}(1988)}]{Fegley1988}
{Fegley}, B., {Prinn}, R.~G., 1988. {Chemical constraints on the water and
  total oxygen abundances in the deep atmosphere of Jupiter}. The Astrophysical
  Journal 324, 621--625.

\bibitem[{{Feuchtgruber} et~al.(2013){Feuchtgruber}, {Lellouch}, {Orton}, {de
  Graauw}, {Vandenbussche}, {Swinyard}, {Moreno}, {Jarchow}, {Billebaud},
  {Cavali{\'e}}, {Sidher}, and {Hartogh}}]{Feuchtgruber2013}
{Feuchtgruber}, H., {Lellouch}, E., {Orton}, G., {de Graauw}, T.,
  {Vandenbussche}, B., {Swinyard}, B., {Moreno}, R., {Jarchow}, C.,
  {Billebaud}, F., {Cavali{\'e}}, T., {Sidher}, S., {Hartogh}, P., 2013. {The
  D/H ratio in the atmospheres of Uranus and Neptune from Herschel-PACS
  observations}. \aap 551, A126.

\bibitem[{{Fletcher} et~al.(2007){Fletcher}, {Irwin}, {Teanby}, {Orton},
  {Parrish}, {Calcutt}, {Bowles}, {de Kok}, {Howett}, and
  {Taylor}}]{Fletcher2007}
{Fletcher}, L.~N., {Irwin}, P.~G.~J., {Teanby}, N.~A., {Orton}, G.~S.,
  {Parrish}, P.~D., {Calcutt}, S.~B., {Bowles}, N., {de Kok}, R., {Howett}, C.,
  {Taylor}, F.~W., 2007. {The meridional phosphine distribution in Saturn's
  upper troposphere from Cassini/CIRS observations}. \icarus 188, 72--88.

\bibitem[{{Fletcher} et~al.(2009){Fletcher}, {Orton}, {Teanby}, {Irwin}, and
  {Bjoraker}}]{Fletcher2009}
{Fletcher}, L.~N., {Orton}, G.~S., {Teanby}, N.~A., {Irwin}, P.~G.~J.,
  {Bjoraker}, G.~L., 2009. {Methane and its isotopologues on Saturn from
  Cassini/CIRS observations}. \icarus 199, 351--367.

\bibitem[{{Fletcher} et~al.(2010){Fletcher}, {Drossart}, {Burgdorf}, {Orton},
  and {Encrenaz}}]{Fletcher2010}
{Fletcher}, L.~N., {Drossart}, P., {Burgdorf}, M., {Orton}, G.~S., {Encrenaz},
  T., 2010. {Neptune's atmospheric composition from AKARI infrared
  spectroscopy}. \aap 514, A17.

\bibitem[{{Fletcher} et~al.(2012){Fletcher}, {Swinyard}, {Salji},
  {Polehampton}, {Fulton}, {Sidher}, {Lellouch}, {Moreno}, {Orton},
  {Cavali{\'e}}, {Courtin}, {Rengel}, {Sagawa}, {Davis}, {Hartogh}, {Naylor},
  {Walker}, and {Lim}}]{Fletcher2012}
{Fletcher}, L.~N., {Swinyard}, B., {Salji}, C., {Polehampton}, E., {Fulton},
  T., {Sidher}, S., {Lellouch}, E., {Moreno}, R., {Orton}, G., {Cavali{\'e}},
  T., {Courtin}, R., {Rengel}, M., {Sagawa}, H., {Davis}, G.~R., {Hartogh}, P.,
  {Naylor}, D., {Walker}, H., {Lim}, T., 2012. {Sub-millimetre spectroscopy of
  Saturn's trace gases from Herschel/SPIRE}. \aap 539, A44.

\bibitem[{{Fletcher} et~al.(2016){Fletcher}, {Irwin}, {Achterberg}, {Orton},
  and {Flasar}}]{Fletcher2016}
{Fletcher}, L.~N., {Irwin}, P.~G.~J., {Achterberg}, R.~K., {Orton}, G.~S.,
  {Flasar}, F.~M., 2016. {Seasonal variability of Saturn's tropospheric
  temperatures, winds and para-H$_{2}$ from Cassini far-IR spectroscopy}.
  \icarus 264, 137--159.

\bibitem[{{Fray} and {Schmitt}(2009)}]{Fray2009}
{Fray}, N., {Schmitt}, B., 2009. {Sublimation of ices of astrophysical
  interest: A bibliographic review}. \planss 57, 2053--2080.

\bibitem[{{Fressin} et~al.(2013){Fressin}, {Torres}, {Charbonneau}, {Bryson},
  {Christiansen}, {Dressing}, {Jenkins}, {Walkowicz}, and
  {Batalha}}]{Fressin2013}
{Fressin}, F., {Torres}, G., {Charbonneau}, D., {Bryson}, S.~T.,
  {Christiansen}, J., {Dressing}, C.~D., {Jenkins}, J.~M., {Walkowicz}, L.~M.,
  {Batalha}, N.~M., 2013. {The False Positive Rate of Kepler and the Occurrence
  of Planets}. \apj 766, 81.

\bibitem[{{Gautier} et~al.(2001){Gautier}, {Hersant}, {Mousis}, and
  {Lunine}}]{Gautier2001}
{Gautier}, D., {Hersant}, F., {Mousis}, O., {Lunine}, J.~I., 2001. {Enrichments
  in Volatiles in Jupiter: A New Interpretation of the Galileo Measurements}.
  \apjl 550, L227--L230.

\bibitem[{{Gautier} and {Hersant}(2005)}]{Gautier2005}
{Gautier}, D., {Hersant}, F., 2005. {Formation and Composition of
  Planetesimals}. \ssr 116, 25--52.

\bibitem[{{Gierasch} and {Conrath}(1985)}]{Gierasch1985}
{Gierasch}, P.~J., {Conrath}, B.~J., 1985. {Energy conversion processes in the
  outer planets}. In: {Hunt}, G.~E. (Ed.), Recent Advances in Planetary
  Meteorology. pp. 121--146.

\bibitem[{{Gomes} et~al.(2005){Gomes}, {Levison}, {Tsiganis}, and
  {Morbidelli}}]{Gomes2005}
{Gomes}, R., {Levison}, H.~F., {Tsiganis}, K., {Morbidelli}, A., 2005. {Origin
  of the cataclysmic Late Heavy Bombardment period of the terrestrial planets}.
  \nat 435, 466--469.

\bibitem[{{Gordon} and {McBride}(1994)}]{Gordon1994}
{Gordon}, S., {McBride}, B.~J., 1994. Computer program for calculation of
  complex chemical equilibrium compositions and applications. part 1: Analysis.
  NASA Reference Publication 1311.

\bibitem[{{Greve} et~al.(1994){Greve}, {Steppe}, {Graham}, and
  {Schalinski}}]{Greve1994}
{Greve}, A., {Steppe}, H., {Graham}, D., {Schalinski}, C.~J., 1994. {Disk
  brightness temperature of the planets at 43 GHz (and 43 GHz flux densities of
  some continuum sources)}. \aap 286, 654--658.

\bibitem[{{Griffin} and {Orton}(1993)}]{Griffin1993}
{Griffin}, M.~J., {Orton}, G.~S., 1993. {The near-millimeter brightness
  temperature spectra of Uranus and Neptune}. \icarus 105, 537.

\bibitem[{{Guillot}(1995)}]{Guillot1995}
{Guillot}, T., 1995. {Condensation of Methane, Ammonia, and Water and the
  Inhibition of Convection in Giant Planets}. Science 269, 1697--1699.

\bibitem[{{Gulkis} et~al.(1978){Gulkis}, {Janssen}, and {Olsen}}]{Gulkis1978}
{Gulkis}, S., {Janssen}, M.~A., {Olsen}, E.~T., 1978. {Evidence for the
  depletion of ammonia in the Uranus atmosphere}. \icarus 34, 10--19.

\bibitem[{{Hanel} et~al.(1981){Hanel}, {Conrath}, {Herath}, {Kunde}, and
  {Pirraglia}}]{Hanel1981}
{Hanel}, R., {Conrath}, B., {Herath}, L., {Kunde}, V., {Pirraglia}, J., 1981.
  {Albedo, internal heat, and energy balance of Jupiter - Preliminary results
  of the Voyager infrared investigation}. \jgr 86, 8705--8712.

\bibitem[{{Hanel} et~al.(1983){Hanel}, {Conrath}, {Kunde}, {Pearl}, and
  {Pirraglia}}]{Hanel1983}
{Hanel}, R.~A., {Conrath}, B.~J., {Kunde}, V.~G., {Pearl}, J.~C., {Pirraglia},
  J.~A., 1983. {Albedo, internal heat flux, and energy balance of Saturn}.
  \icarus 53, 262--285.

\bibitem[{{Held} and {Dryer}(1998)}]{Held1998}
{Held}, T.~J., {Dryer}, F.~L., 1998. {A comprehensive mechanism for methanol
  oxidation}. International Journal of Chemical Kinetics 30, 805--830.

\bibitem[{{Helled} et~al.(2011){Helled}, {Anderson}, {Podolak}, and
  {Schubert}}]{Helled2011}
{Helled}, R., {Anderson}, J.~D., {Podolak}, M., {Schubert}, G., 2011. {Interior
  Models of Uranus and Neptune}. \apj 726, 15.

\bibitem[{{Helled} and {Lunine}(2014)}]{Helled2014}
{Helled}, R., {Lunine}, J., 2014. {Measuring Jupiter's water abundance by Juno:
  the link between interior and formation models}. \mnras 441, 2273--2279.

\bibitem[{{Hersant} et~al.(2001){Hersant}, {Gautier}, and
  {Hur{\'e}}}]{Hersant2001}
{Hersant}, F., {Gautier}, D., {Hur{\'e}}, J.-M., 2001. {A Two-dimensional Model
  for the Primordial Nebula Constrained by D/H Measurements in the Solar
  System: Implications for the Formation of Giant Planets}. \apj 554, 391--407.

\bibitem[{{Hersant} et~al.(2004){Hersant}, {Gautier}, and
  {Lunine}}]{Hersant2004}
{Hersant}, F., {Gautier}, D., {Lunine}, J.~I., 2004. {Enrichment in volatiles
  in the giant planets of the Solar System}. \planss 52, 623--641.

\bibitem[{{Hersant} et~al.(2008){Hersant}, {Gautier}, {Tobie}, and
  {Lunine}}]{Hersant2008}
{Hersant}, F., {Gautier}, D., {Tobie}, G., {Lunine}, J.~I., 2008.
  {Interpretation of the carbon abundance in Saturn measured by Cassini}.
  \planss 56, 1103--1111.

\bibitem[{{Hidaka} et~al.(1989){Hidaka}, {Oki}, {Kawano}, and
  {Higashihara}}]{Hidaka1989}
{Hidaka}, Y., {Oki}, T., {Kawano}, H., {Higashihara}, T., 1989. {Thermal
  decomposition of methanol in shock waves}. \jchemphys 93, 7134--7139.

\bibitem[{{Hofstadter}(1993)}]{Hofstadter1993}
{Hofstadter}, M.~D., 1993. {Microwave Imaging of Neptune's Troposphere}.
  In: AAS/Division for Planetary Sciences Meeting Abstracts \#25. Vol.~25 of
  Bulletin of the American Astronomical Society. p. 1077.

\bibitem[{{Hofstadter} and {Butler}(2003)}]{Hofstadter2003}
{Hofstadter}, M.~D., {Butler}, B.~J., 2003. {Seasonal change in the deep
  atmosphere of Uranus}. \icarus 165, 168--180.

\bibitem[{{Karkoschka} and {Tomasko}(2009)}]{Karkoschka2009}
{Karkoschka}, E., {Tomasko}, M., 2009. {The haze and methane distributions on
  Uranus from HST-STIS spectroscopy}. \icarus 202, 287--309.

\bibitem[{{Karkoschka} and {Tomasko}(2011)}]{Karkoschka2011}
{Karkoschka}, E., {Tomasko}, M.~G., 2011. {The haze and methane distributions
  on Neptune from HST-STIS spectroscopy}. \icarus 211, 780--797.

\bibitem[{{Karpowicz} and {Steffes}(2013)}]{Karpowicz2013}
{Karpowicz}, B.~M., {Steffes}, P.~G., 2013. {Investigating the
  H$_{2}$-He-H$_{2}$O-CH$_{4}$ equation of state in the deep troposphere of
  Jupiter}. \icarus 223, 277--297.

\bibitem[{{Klein} and {Hofstadter}(2006)}]{Klein2006}
{Klein}, M.~J., {Hofstadter}, M.~D., 2006. {Long-term variations in the
  microwave brightness temperature of the Uranus atmosphere}. \icarus 184,
  170--180.

\bibitem[{{Leconte} and {Chabrier}(2012)}]{Leconte2012}
{Leconte}, J., {Chabrier}, G., 2012. {A new vision of giant planet interiors:
  Impact of double diffusive convection}. \aap 540, A20.

\bibitem[{{Leconte} et~al.(2013){Leconte}, {Forget}, {Charnay}, {Wordsworth},
  and {Pottier}}]{Leconte2013}
{Leconte}, J., {Forget}, F., {Charnay}, B., {Wordsworth}, R., {Pottier}, A.,
  2013. {Increased insolation threshold for runaway greenhouse processes on
  Earth-like planets}. \nat 504, 268--271.

\bibitem[{{Leconte} et~al.(2017){Leconte}, {Selsis}, {Hersant}, and
  {Guillot}}]{Leconte2017}
{Leconte}, J., {Selsis}, F., {Hersant}, F., {Guillot}, T., 2016.
  {Condensation-inhibited convection in hydrogen-rich atmospheres. Stability
  against double-diffusive processes and thermal profiles for Jupiter, Saturn,
  Uranus, and Neptune}. \aap, 598, A98.

\bibitem[{{Ledoux}(1947)}]{Ledoux1947}
{Ledoux}, P., 1947. {Stellar Models with Convection and with Discontinuity of
  the Mean Molecular Weight}. \apj 105, 305.

\bibitem[{{Lellouch} et~al.(2001){Lellouch}, {B{\'e}zard}, {Fouchet},
  {Feuchtgruber}, {Encrenaz}, and {de Graauw}}]{Lellouch2001}
{Lellouch}, E., {B{\'e}zard}, B., {Fouchet}, T., {Feuchtgruber}, H.,
  {Encrenaz}, T., {de Graauw}, T., 2001. {The deuterium abundance in Jupiter
  and Saturn from ISO-SWS observations}. \aap 370, 610--622.

\bibitem[{{Lellouch} et~al.(2005){Lellouch}, {Moreno}, and
  {Paubert}}]{Lellouch2005}
{Lellouch}, E., {Moreno}, R., {Paubert}, G., 2005. {A dual origin for Neptune's
  carbon monoxide?} \aap 430, L37--L40.

\bibitem[{{Lellouch} et~al.(2010){Lellouch}, {Hartogh}, {Feuchtgruber},
  {Vandenbussche}, {de Graauw}, {Moreno}, {Jarchow}, {Cavali{\'e}}, {Orton},
  {Banaszkiewicz}, {Blecka}, {Bockel{\'e}e-Morvan}, {Crovisier}, {Encrenaz},
  {Fulton}, {K{\"u}ppers}, {Lara}, {Lis}, {Medvedev}, {Rengel}, {Sagawa},
  {Swinyard}, {Szutowicz}, {Bensch}, {Bergin}, {Billebaud}, {Biver}, {Blake},
  {Blommaert}, {Cernicharo}, {Courtin}, {Davis}, {Decin}, {Encrenaz},
  {Gonzalez}, {Jehin}, {Kidger}, {Naylor}, {Portyankina}, {Schieder}, {Sidher},
  {Thomas}, {de Val-Borro}, {Verdugo}, {Waelkens}, {Walker}, {Aarts}, {Comito},
  {Kawamura}, {Maestrini}, {Peacocke}, {Teipen}, {Tils}, and
  {Wildeman}}]{Lellouch2010}
{Lellouch}, E., {Hartogh}, P., {Feuchtgruber}, H., {Vandenbussche}, B., {de
  Graauw}, T., {Moreno}, R., {Jarchow}, C., {Cavali{\'e}}, T., {Orton}, G.,
  {Banaszkiewicz}, M., {Blecka}, M.~I., {Bockel{\'e}e-Morvan}, D., {Crovisier},
  J., {Encrenaz}, T., {Fulton}, T., {K{\"u}ppers}, M., {Lara}, L.~M., {Lis},
  D.~C., {Medvedev}, A.~S., {Rengel}, M., {Sagawa}, H., {Swinyard}, B.,
  {Szutowicz}, S., {Bensch}, F., {Bergin}, E., {Billebaud}, F., {Biver}, N.,
  {Blake}, G.~A., {Blommaert}, J.~A.~D.~L., {Cernicharo}, J., {Courtin}, R.,
  {Davis}, G.~R., {Decin}, L., {Encrenaz}, P., {Gonzalez}, A., {Jehin}, E.,
  {Kidger}, M., {Naylor}, D., {Portyankina}, G., {Schieder}, R., {Sidher}, S.,
  {Thomas}, N., {de Val-Borro}, M., {Verdugo}, E., {Waelkens}, C., {Walker},
  H., {Aarts}, H., {Comito}, C., {Kawamura}, J.~H., {Maestrini}, A.,
  {Peacocke}, T., {Teipen}, R., {Tils}, T., {Wildeman}, K., 2010. {First
  results of Herschel-PACS observations of Neptune}. \aap 518, L152.

\bibitem[{{Lellouch} et~al.(2015){Lellouch}, {Moreno}, {Orton}, {Feuchtgruber},
  {Cavali{\'e}}, {Moses}, {Hartogh}, {Jarchow}, and {Sagawa}}]{Lellouch2015}
{Lellouch}, E., {Moreno}, R., {Orton}, G.~S., {Feuchtgruber}, H.,
  {Cavali{\'e}}, T., {Moses}, J.~I., {Hartogh}, P., {Jarchow}, C., {Sagawa},
  H., 2015. {New constraints on the CH$_{4}$ vertical profile in Uranus
  and Neptune from Herschel observations}. \aap 579, A121.

\bibitem[{{Liebe} and {Dillon}(1969)}]{Liebe1969}
{Liebe}, H.~J., {Dillon}, T.~A., 1969. {Accurate Foreign?Gas?Broadening
  Parameters of the 22?GHz H2O Line from Refraction Spectroscopy}. The Journal
  of Chemical Physics 50, 727--732.

\bibitem[{{Lindal} et~al.(1987){Lindal}, {Lyons}, {Sweetnam}, {Eshleman}, and
  {Hinson}}]{Lindal1987}
{Lindal}, G.~F., {Lyons}, J.~R., {Sweetnam}, D.~N., {Eshleman}, V.~R.,
  {Hinson}, D.~P., 1987. {The atmosphere of Uranus - Results of radio
  occultation measurements with Voyager 2}. \jgr 92, 14987--15001.

\bibitem[{{Lindal} et~al.(1990){Lindal}, {Lyons}, {Sweetnam}, {Eshleman}, and
  {Hinson}}]{Lindal1990}
{Lindal}, G.~F., {Lyons}, J.~R., {Sweetnam}, D.~N., {Eshleman}, V.~R.,
  {Hinson}, D.~P., 1990. {The atmosphere of Neptune - Results of radio
  occultation measurements with the Voyager 2 spacecraft}. \grl 17, 1733--1736.

\bibitem[{{Lindal}(1992)}]{Lindal1992}
{Lindal}, G.~F., 1992. {The atmosphere of Neptune - an analysis of radio
  occultation data acquired with Voyager 2}. \aj 103, 967--982.

\bibitem[{{Lodders} and {Fegley}(1994)}]{Lodders1994}
{Lodders}, K., {Fegley}, Jr., B., 1994. {The origin of carbon monoxide in
  Neptunes's atmosphere}. \icarus 112, 368--375.

\bibitem[{{Lodders}(2004)}]{Lodders2004}
{Lodders}, K., 1994. {Jupiter Formed with More Tar 
  than Ice}. \apj 611, 587--597.

\bibitem[{{Lodders}(2010)}]{Lodders2010}
{Lodders}, K., 2010. {Solar System Abundances of the Elements}. In: {Goswami},
  A., {Reddy}, B.~E. (Eds.), Principles and Perspectives in Cosmochemistry. p.
  379.

\bibitem[{{Lunine} and {Stevenson}(1985)}]{Lunine1985}
{Lunine}, J.~I., {Stevenson}, D.~J., 1985. {Thermodynamics of clathrate hydrate
  at low and high pressures with application to the outer solar system}. \apjs
  58, 493--531.

\bibitem[{{Luszcz-Cook} and {de Pater}(2013)}]{Luszcz-Cook2013}
{Luszcz-Cook}, S.~H., {de Pater}, I., 2013. {Constraining the origins of
  Neptune's carbon monoxide abundance with CARMA millimeter-wave observations}.
  \icarus 222, 379--400.

\bibitem[{{Mahaffy} et~al.(2000){Mahaffy}, {Niemann}, {Alpert}, {Atreya},
  {Demick}, {Donahue}, {Harpold}, and {Owen}}]{Mahaffy2000}
{Mahaffy}, P.~R., {Niemann}, H.~B., {Alpert}, A., {Atreya}, S.~K., {Demick},
  J., {Donahue}, T.~M., {Harpold}, D.~N., {Owen}, T.~C., 2000. {Noble gas
  abundance and isotope ratios in the atmosphere of Jupiter from the Galileo
  Probe Mass Spectrometer}. \jgr 105, 15061--15072.

\bibitem[{{Mandt} et~al.(2015){Mandt}, {Mousis}, {Marty}, {Cavali{\'e}},
  {Harris}, {Hartogh}, and {Willacy}}]{Mandt2015}
{Mandt}, K.~E., {Mousis}, O., {Marty}, B., {Cavali{\'e}}, T., {Harris}, W.,
  {Hartogh}, P., {Willacy}, K., 2015. {Constraints from Comets on the
  Formation and Volatile Acquisition of the Planets and Satellites}. \ssr 197,
  297--342.

\bibitem[{{Marten} et~al.(1993){Marten}, {Gautier}, {Owen}, {Sanders},
  {Matthews}, {Atreya}, {Tilanus}, and {Deane}}]{Marten1993}
{Marten}, A., {Gautier}, D., {Owen}, T., {Sanders}, D.~B., {Matthews}, H.~E.,
  {Atreya}, S.~K., {Tilanus}, R.~P.~J., {Deane}, J.~R., 1993. {First
  observations of CO and HCN on Neptune and Uranus at millimeter wavelengths
  and the implications for atmospheric chemistry}. \apj 406, 285--297.

\bibitem[{{Marten} et~al.(2005){Marten}, {Matthews}, {Owen}, {Moreno},
  {Hidayat}, and {Biraud}}]{Marten2005}
{Marten}, A., {Matthews}, H.~E., {Owen}, T., {Moreno}, R., {Hidayat}, T.,
  {Biraud}, Y., 2005. {Improved constraints on Neptune's atmosphere from
  submillimetre-wavelength observations}. \aap 429, 1097--1105.

\bibitem[{{Masters} et~al.(2014){Masters}, {Achilleos}, {Agnor}, {Campagnola},
  {Charnoz}, {Christophe}, {Coates}, {Fletcher}, {Jones}, {Lamy}, {Marzari},
  {Nettelmann}, {Ruiz}, {Ambrosi}, {Andre}, {Bhardwaj}, {Fortney}, {Hansen},
  {Helled}, {Moragas-Klostermeyer}, {Orton}, {Ray}, {Reynaud}, {Sergis},
  {Srama}, and {Volwerk}}]{Masters2014}
{Masters}, A., {Achilleos}, N., {Agnor}, C.~B., {Campagnola}, S., {Charnoz},
  S., {Christophe}, B., {Coates}, A.~J., {Fletcher}, L.~N., {Jones}, G.~H.,
  {Lamy}, L., {Marzari}, F., {Nettelmann}, N., {Ruiz}, J., {Ambrosi}, R.,
  {Andre}, N., {Bhardwaj}, A., {Fortney}, J.~J., {Hansen}, C.~J., {Helled}, R.,
  {Moragas-Klostermeyer}, G., {Orton}, G., {Ray}, L., {Reynaud}, S., {Sergis},
  N., {Srama}, R., {Volwerk}, M., 2014. {Neptune and Triton: Essential
  pieces of the Solar System puzzle}. \planss 104, 108--121.

\bibitem[{{Matousek}(2007)}]{Matousek2007}
{Matousek}, S., 2007. {The Juno New Frontiers mission}. Acta Astronautica 61,
  932--939.

\bibitem[{{Morbidelli} et~al.(2005){Morbidelli}, {Levison}, {Tsiganis}, and
  {Gomes}}]{Morbidelli2005}
{Morbidelli}, A., {Levison}, H.~F., {Tsiganis}, K., {Gomes}, R., 2005. {Chaotic
  capture of Jupiter's Trojan asteroids in the early Solar System}. \nat 435,
  462--465.

\bibitem[{{Moreno}(1998)}]{Moreno1998}
{Moreno}, R.~M., 1998. {Observations millim\'etriques et submillim\'etriques
  des plan\`etes g\'eantes. \'Etude de Jupiter apr\`es la chute de la com\`ete
  SL9.} Ph.D. thesis, Universit\'e Paris VI.

\bibitem[{{Moses} et~al.(2011){Moses}, {Visscher}, {Fortney}, {Showman},
  {Lewis}, {Griffith}, {Klippenstein}, {Shabram}, {Friedson}, {Marley}, and
  {Freedman}}]{Moses2011}
{Moses}, J.~I., {Visscher}, C., {Fortney}, J.~J., {Showman}, A.~P., {Lewis},
  N.~K., {Griffith}, C.~A., {Klippenstein}, S.~J., {Shabram}, M., {Friedson},
  A.~J., {Marley}, M.~S., {Freedman}, R.~S., 2011. {Disequilibrium Carbon,
  Oxygen, and Nitrogen Chemistry in the Atmospheres of HD 189733b and HD
  209458b}. \apj 737, 15.

\bibitem[{{Moses} et~al.(2013){Moses}, {Madhusudhan}, {Visscher}, and
  {Freedman}}]{Moses2013}
{Moses}, J.~I., {Madhusudhan}, N., {Visscher}, C., {Freedman}, R.~S., 2013.
  {Chemical Consequences of the C/O Ratio on Hot Jupiters: Examples from
  WASP-12b, CoRoT-2b, XO-1b, and HD 189733b}. \apj 763, 25.

\bibitem[{{Moses}(2014)}]{Moses2014}
{Moses}, J.~I., 2014. {Chemical kinetics on extrasolar planets}.
  Philosophical Transactions of the Royal Society of London Series A 372,
  20130073--20130073.

\bibitem[{{Mousis} et~al.(2012){Mousis}, {Lunine}, {Madhusudhan}, and
  {Johnson}}]{Mousis2012}
{Mousis}, O., {Lunine}, J.~I., {Madhusudhan}, N., {Johnson}, T.~V., 2012.
  {Nebular Water Depletion as the Cause of Jupiter's Low Oxygen Abundance}.
  \apjl 751, L7.

\bibitem[{{Mousis} et~al.(2014){Mousis}, {Fletcher}, {Lebreton}, {Wurz},
  {Cavali{\'e}}, {Coustenis}, {Courtin}, {Gautier}, {Helled}, {Irwin}, {Morse},
  {Nettelmann}, {Marty}, {Rousselot}, {Venot}, {Atkinson}, {Waite}, {Reh},
  {Simon}, {Atreya}, {Andr{\'e}}, {Blanc}, {Daglis}, {Fischer}, {Geppert},
  {Guillot}, {Hedman}, {Hueso}, {Lellouch}, {Lunine}, {Murray}, {O`Donoghue},
  {Rengel}, {S{\'a}nchez-Lavega}, {Schmider}, {Spiga}, {Spilker}, {Petit},
  {Tiscareno}, {Ali-Dib}, {Altwegg}, {Bolton}, {Bouquet}, {Briois}, {Fouchet},
  {Guerlet}, {Kostiuk}, {Lebleu}, {Moreno}, {Orton}, and {Poncy}}]{Mousis2014}
{Mousis}, O., {Fletcher}, L.~N., {Lebreton}, J.-P., {Wurz}, P., {Cavali{\'e}},
  T., {Coustenis}, A., {Courtin}, R., {Gautier}, D., {Helled}, R., {Irwin},
  P.~G.~J., {Morse}, A.~D., {Nettelmann}, N., {Marty}, B., {Rousselot}, P.,
  {Venot}, O., {Atkinson}, D.~H., {Waite}, J.~H., {Reh}, K.~R., {Simon}, A.~A.,
  {Atreya}, S., {Andr{\'e}}, N., {Blanc}, M., {Daglis}, I.~A., {Fischer}, G.,
  {Geppert}, W.~D., {Guillot}, T., {Hedman}, M.~M., {Hueso}, R., {Lellouch},
  E., {Lunine}, J.~I., {Murray}, C.~D., {O`Donoghue}, J., {Rengel}, M.,
  {S{\'a}nchez-Lavega}, A., {Schmider}, F.-X., {Spiga}, A., {Spilker}, T.,
  {Petit}, J.-M., {Tiscareno}, M.~S., {Ali-Dib}, M., {Altwegg}, K., {Bolton},
  S.~J., {Bouquet}, A., {Briois}, C., {Fouchet}, T., {Guerlet}, S., {Kostiuk},
  T., {Lebleu}, D., {Moreno}, R., {Orton}, G.~S., {Poncy}, J., 2014.
  {Scientific rationale for Saturn's in situ exploration}. \planss 104, 29--47.

\bibitem[{{Mousis} et~al.(2016){Mousis}, {Atkinson}, {Spilker}, {Venkatapathy},
  {Poncy}, {Frampton}, {Coustenis}, {Reh}, {Lebreton}, {Fletcher}, {Hueso},
  {Amato}, {Colaprete}, {Ferri}, {Stam}, {Wurz}, {Atreya}, {Aslam}, {Banfield},
  {Calcutt}, {Fischer}, {Holland}, {Keller}, {Kessler}, {Leese}, {Levacher},
  {Morse}, {Mu{\~n}oz}, {Renard}, {Sheridan}, {Schmider}, {Snik}, {Waite},
  {Bird}, {Cavali{\'e}}, {Deleuil}, {Fortney}, {Gautier}, {Guillot}, {Lunine},
  {Marty}, {Nixon}, {Orton}, and {S{\'a}nchez-Lavega}}]{Mousis2016}
{Mousis}, O., {Atkinson}, D.~H., {Spilker}, T., {Venkatapathy}, E., {Poncy},
  J., {Frampton}, R., {Coustenis}, A., {Reh}, K., {Lebreton}, J.-P.,
  {Fletcher}, L.~N., {Hueso}, R., {Amato}, M.~J., {Colaprete}, A., {Ferri}, F.,
  {Stam}, D., {Wurz}, P., {Atreya}, S., {Aslam}, S., {Banfield}, D.~J.,
  {Calcutt}, S., {Fischer}, G., {Holland}, A., {Keller}, C., {Kessler}, E.,
  {Leese}, M., {Levacher}, P., {Morse}, A., {Mu{\~n}oz}, O., {Renard}, J.-B.,
  {Sheridan}, S., {Schmider}, F.-X., {Snik}, F., {Waite}, J.~H., {Bird}, M.,
  {Cavali{\'e}}, T., {Deleuil}, M., {Fortney}, J., {Gautier}, D., {Guillot},
  T., {Lunine}, J.~I., {Marty}, B., {Nixon}, C., {Orton}, G.~S.,
  {S{\'a}nchez-Lavega}, A., 2016. {The Hera Saturn entry probe mission}.
  \planss 130, 80--103.

\bibitem[{{Muhleman} and {Berge}(1991)}]{Muhleman1991}
{Muhleman}, D.~O., {Berge}, G.~L., 1991. {Observations of Mars, Uranus,
  Neptune, Io, Europa, Ganymede, and Callisto at a wavelength of 2.66 MM}.
  \icarus 92, 263--272.

\bibitem[{{Nettelmann} et~al.(2008){Nettelmann}, {Holst}, {Kietzmann},
  {French}, {Redmer}, and {Blaschke}}]{Nettelmann2008}
{Nettelmann}, N., {Holst}, B., {Kietzmann}, A., {French}, M., {Redmer}, R.,
  {Blaschke}, D., 2008. {Ab Initio Equation of State Data for Hydrogen, Helium,
  and Water and the Internal Structure of Jupiter}. \apj 683, 1217--1228.

\bibitem[{{Nettelmann} et~al.(2013){Nettelmann}, {Helled}, {Fortney}, and
  {Redmer}}]{Nettelmann2013}
{Nettelmann}, N., {Helled}, R., {Fortney}, J.~J., {Redmer}, R., 2013. {New
  indication for a dichotomy in the interior structure of Uranus and Neptune
  from the application of modified shape and rotation data}. \planss 77,
  143--151.

\bibitem[{{Nettelmann} et~al.(2016){Nettelmann}, {Wang}, {Fortney}, {Hamel},
  {Yellamilli}, {Bethkenhagen}, and {Redmer}}]{Nettelmann2016}
{Nettelmann}, N., {Wang}, K., {Fortney}, J.~J., {Hamel}, S., {Yellamilli}, S.,
  {Bethkenhagen}, M., {Redmer}, R., 2016. {Uranus evolution models with
  simple thermal boundary layers}. \icarus 275, 107--116.

\bibitem[{{Niemann} et~al.(1998){Niemann}, {Atreya}, {Carignan}, {Donahue},
  {Haberman}, {Harpold}, {Hartle}, {Hunten}, {Kasprzak}, {Mahaffy}, {Owen}, and
  {Way}}]{Niemann1998}
{Niemann}, H.~B., {Atreya}, S.~K., {Carignan}, G.~R., {Donahue}, T.~M.,
  {Haberman}, J.~A., {Harpold}, D.~N., {Hartle}, R.~E., {Hunten}, D.~M.,
  {Kasprzak}, W.~T., {Mahaffy}, P.~R., {Owen}, T.~C., {Way}, S.~H., 1998. {The
  composition of the Jovian atmosphere as determined by the Galileo probe mass
  spectrometer}. \jgr 103, 22831--22846.

\bibitem[{{Norwood} et~al.(2016{\natexlab{a}}){Norwood}, {Hammel}, {Milam},
  {Stansberry}, {Lunine}, {Chanover}, {Hines}, {Sonneborn}, {Tiscareno},
  {Brown}, and {Ferruit}}]{Norwood2016b}
{Norwood}, J., {Hammel}, H., {Milam}, S., {Stansberry}, J., {Lunine}, J.,
  {Chanover}, N., {Hines}, D., {Sonneborn}, G., {Tiscareno}, M., {Brown}, M.,
  {Ferruit}, P., 2016{\natexlab{a}}. {Solar System Observations with the
  James Webb Space Telescope}. \pasp 128~(2), 025004.

\bibitem[{{Norwood} et~al.(2016{\natexlab{b}}){Norwood}, {Moses}, {Fletcher},
  {Orton}, {Irwin}, {Atreya}, {Rages}, {Cavali{\'e}}, {S{\'a}nchez-Lavega},
  {Hueso}, and {Chanover}}]{Norwood2016a}
{Norwood}, J., {Moses}, J., {Fletcher}, L.~N., {Orton}, G., {Irwin}, P.~G.~J.,
  {Atreya}, S., {Rages}, K., {Cavali{\'e}}, T., {S{\'a}nchez-Lavega}, A.,
  {Hueso}, R., {Chanover}, N., 2016{\natexlab{b}}. {Giant Planet
  Observations with the James Webb Space Telescope}. \pasp 128~(1), 018005.

\bibitem[{{Orton} et~al.(1986){Orton}, {Griffin}, {Ade}, {Nolt}, and
  {Radostitz}}]{Orton1986}
{Orton}, G.~S., {Griffin}, M.~J., {Ade}, P.~A.~R., {Nolt}, I.~G., {Radostitz},
  J.~V., 1986. {Submillimeter and millimeter observations of Uranus and
  Neptune}. \icarus 67, 289--304.

\bibitem[{{Orton} et~al.(2014){Orton}, {Fletcher}, {Moses}, {Mainzer}, {Hines},
  {Hammel}, {Martin-Torres}, {Burgdorf}, {Merlet}, and {Line}}]{Orton2014a}
{Orton}, G.~S., {Fletcher}, L.~N., {Moses}, J.~I., {Mainzer}, A.~K., {Hines},
  D., {Hammel}, H.~B., {Martin-Torres}, F.~J., {Burgdorf}, M., {Merlet}, C.,
  {Line}, M.~R., 2014. {Mid-infrared spectroscopy of Uranus from the
  Spitzer Infrared Spectrometer: 1. Determination of the mean temperature
  structure of the upper troposphere and stratosphere}. \icarus 243, 494--513.

\bibitem[{{Owen} et~al.(1999){Owen}, {Mahaffy}, {Niemann}, {Atreya}, {Donahue},
  {Bar-Nun}, and {de Pater}}]{Owen1999}
{Owen}, T., {Mahaffy}, P., {Niemann}, H.~B., {Atreya}, S., {Donahue}, T.,
  {Bar-Nun}, A., {de Pater}, I., 1999. {A low-temperature origin for the
  planetesimals that formed Jupiter}. \nat 402, 269--270.

\bibitem[{{Owen} and {Encrenaz}(2003)}]{Owen2003}
{Owen}, T., {Encrenaz}, T., 2003. {Element Abundances and Isotope Ratios in the
  Giant Planets and Titan}. \ssr 106, 121--138.

\bibitem[{{Owen} and {Encrenaz}(2006)}]{Owen2006}
{Owen}, T., {Encrenaz}, T., 2006. {Compositional constraints on giant planet
  formation}. \planss 54, 1188--1196.

\bibitem[{{Owen}(2007)}]{Owen2007}
{Owen}, T.~C., 2007. {Planetary Atmospheres}. \ssr 130, 97--104.

\bibitem[{{Pearl} et~al.(1990){Pearl}, {Conrath}, {Hanel}, and
  {Pirraglia}}]{Pearl1990}
{Pearl}, J.~C., {Conrath}, B.~J., {Hanel}, R.~A., {Pirraglia}, J.~A., 1990.
  {The albedo, effective temperature, and energy balance of Uranus, as
  determined from Voyager IRIS data}. \icarus 84, 12--28.

\bibitem[{{Pearl} and {Conrath}(1991)}]{Pearl1991}
{Pearl}, J.~C., {Conrath}, B.~J., 1991. {The albedo, effective temperature, and
  energy balance of Neptune, as determined from Voyager data}. \jgr 96, 18921.

\bibitem[{{Pickett} et~al.(1998){Pickett}, {Poynter}, {Cohen}, {Delitsky},
  {Pearson}, and {M{\"u}ller}}]{Pickett1998}
{Pickett}, H.~M., {Poynter}, R.~L., {Cohen}, E.~A., {Delitsky}, M.~L.,
  {Pearson}, J.~C., {M{\"u}ller}, H.~S.~P., 1998. {Submillimeter,
  millimeter and microwave spectral line catalog.} \jqsrt 60, 883--890.

\bibitem[{{Pollack} et~al.(1996){Pollack}, {Hubickyj}, {Bodenheimer},
  {Lissauer}, {Podolak}, and {Greenzweig}}]{Pollack1996}
{Pollack}, J.~B., {Hubickyj}, O., {Bodenheimer}, P., {Lissauer}, J.~J.,
  {Podolak}, M., {Greenzweig}, Y., 1996. {Formation of the Giant Planets by
  Concurrent Accretion of Solids and Gas}. \icarus 124, 62--85.

\bibitem[{{Prinn} and {Barshay}(1977)}]{Prinn1977}
{Prinn}, R.~G., {Barshay}, S.~S., 1977. {Carbon monoxide on Jupiter and
  implications for atmospheric convection}. Science 198, 1031--1034.

\bibitem[{{Romani} et~al.(1989){Romani}, {de Pater}, and {Atreya}}]{Romani1989}
{Romani}, P.~N., {de Pater}, I., {Atreya}, S.~K., 1989. {Neptune's deep
  atmosphere revealed}. \grl 16, 933--936.

\bibitem[{{Rosenqvist} et~al.(1992){Rosenqvist}, {Lellouch}, {Romani},
  {Paubert}, and {Encrenaz}}]{Rosenqvist1992}
{Rosenqvist}, J., {Lellouch}, E., {Romani}, P.~N., {Paubert}, G., {Encrenaz},
  T., 1992. {Millimeter-wave observations of Saturn, Uranus, and Neptune - CO
  and HCN on Neptune}. \apjl 392, L99--L102.

\bibitem[{{Sakashita} and {Hayashi}(1959)}]{Sakashita1959}
{Sakashita}, S., {Hayashi}, C., 1959. {Internal Structure and Evolution of Very
  Massive Stars}. Progress of Theoretical Physics 22, 830--834.

\bibitem[{{Schwarzschild} and {H{\"a}rm}(1958)}]{Schwarzschild1958}
{Schwarzschild}, M., {H{\"a}rm}, R., 1958. {Evolution of Very Massive
  Stars.} \apj 128, 348.

\bibitem[{{Seiff} et~al.(1998){Seiff}, {Kirk}, {Knight}, {Young}, {Mihalov},
  {Young}, {Milos}, {Schubert}, {Blanchard}, and {Atkinson}}]{Seiff1998}
{Seiff}, A., {Kirk}, D.~B., {Knight}, T.~C.~D., {Young}, R.~E., {Mihalov},
  J.~D., {Young}, L.~A., {Milos}, F.~S., {Schubert}, G., {Blanchard}, R.~C.,
  {Atkinson}, D., 1998. {Thermal structure of Jupiter's atmosphere near
  the edge of a 5-{$\mu$}m hot spot in the north equatorial belt}. \jgr 103,
  22857--22890.

\bibitem[{{Smith}(1998)}]{Smith1998}
{Smith}, M.~D., 1998. {Estimation of a Length Scale to Use with the Quench
  Level Approximation for Obtaining Chemical Abundances}. \icarus 132,
  176--184.

\bibitem[{{Sromovsky} and {Fry}(2008)}]{Sromovsky2008}
{Sromovsky}, L.~A., {Fry}, P.~M., 2008. {The methane abundance and structure of
  Uranus' cloud bands inferred from spatially resolved 2006 Keck grism
  spectra}. \icarus 193, 252--266.

\bibitem[{{Sromovsky} et~al.(2011){Sromovsky}, {Fry}, and
  {Kim}}]{Sromovsky2011}
{Sromovsky}, L.~A., {Fry}, P.~M., {Kim}, J.~H., 2011. {Methane on Uranus:
  The case for a compact CH $_{4}$ cloud layer at low latitudes and a severe CH
  $_{4}$ depletion at high-latitudes based on re-analysis of Voyager
  occultation measurements and STIS spectroscopy}. \icarus 215, 292--312.

\bibitem[{{Sromovsky} et~al.(2014){Sromovsky}, {Karkoschka}, {Fry}, {Hammel},
  {de Pater}, and {Rages}}]{Sromovsky2014}
{Sromovsky}, L.~A., {Karkoschka}, E., {Fry}, P.~M., {Hammel}, H.~B., {de
  Pater}, I., {Rages}, K., 2014. {Methane depletion in both polar regions of
  Uranus inferred from HST/STIS and Keck/NIRC2 observations}. \icarus 238,
  137--155.

\bibitem[{{Stone}(1976)}]{Stone1976}
{Stone}, P.~H., 1976. {The meteorology of the Jovian atmosphere}. In:
  {Gehrels}, T. (Ed.), IAU Colloq. 30: Jupiter: Studies of the Interior, Atmosp
  here, Magnetosphere and Satellites. pp. 586--618.

\bibitem[{Teanby and Irwin(2013)}]{Teanby2013}
Teanby, N.~A., Irwin, P. G.~J., 2013. An external origin for carbon monoxide on
  uranus from herschel/spire? \apjl 775~(2), L49.

\bibitem[{{Tsiganis} et~al.(2005){Tsiganis}, {Gomes}, {Morbidelli}, and
  {Levison}}]{Tsiganis2005}
{Tsiganis}, K., {Gomes}, R., {Morbidelli}, A., {Levison}, H.~F., 2005. {Origin
  of the orbital architecture of the giant planets of the Solar System}. \nat
  435, 459--461.

\bibitem[{{Turrini} et~al.(2014){Turrini}, {Politi}, {Peron}, {Grassi},
  {Plainaki}, {Barbieri}, {Lucchesi}, {Magni}, {Altieri}, {Cottini}, {Gorius},
  {Gaulme}, {Schmider}, {Adriani}, and {Piccioni}}]{Turrini2014}
{Turrini}, D., {Politi}, R., {Peron}, R., {Grassi}, D., {Plainaki}, C.,
  {Barbieri}, M., {Lucchesi}, D.~M., {Magni}, G., {Altieri}, F., {Cottini}, V.,
  {Gorius}, N., {Gaulme}, P., {Schmider}, F.-X., {Adriani}, A., {Piccioni}, G.,
  2014. {The comparative exploration of the ice giant planets with twin
  spacecraft: Unveiling the history of our Solar System}. \planss 104, 93--107.

\bibitem[{{Valencia} et~al.(2013){Valencia}, {Guillot}, {Parmentier}, and
  {Freedman}}]{Valencia2013}
{Valencia}, D., {Guillot}, T., {Parmentier}, V., {Freedman}, R.~S., 2013. {Bulk
  Composition of GJ 1214b and Other Sub-Neptune Exoplanets}. \apj 775, 10.

\bibitem[{{Vazan} et~al.(2015){Vazan}, {Helled}, {Kovetz}, and
  {Podolak}}]{Vazan2015}
{Vazan}, A., {Helled}, R., {Kovetz}, A., {Podolak}, M., 2015. {Convection
  and Mixing in Giant Planet Evolution}. \apj 803, 32.

\bibitem[{{Venot} et~al.(2012){Venot}, {H{\'e}brard}, {Ag{\'u}ndez},
  {Dobrijevic}, {Selsis}, {Hersant}, {Iro}, and {Bounaceur}}]{Venot2012}
{Venot}, O., {H{\'e}brard}, E., {Ag{\'u}ndez}, M., {Dobrijevic}, M., {Selsis},
  F., {Hersant}, F., {Iro}, N., {Bounaceur}, R., 2012. {A chemical model for
  the atmosphere of hot Jupiters}. \aap 546, A43.

\bibitem[{{Venot} et~al.(2014){Venot}, {Ag{\'u}ndez}, {Selsis}, {Tessenyi}, and
  {Iro}}]{Venot2014}
{Venot}, O., {Ag{\'u}ndez}, M., {Selsis}, F., {Tessenyi}, M., {Iro}, N., 
  2014. {The atmospheric chemistry of the warm Neptune GJ 3470b: Influence of
  metallicity and temperature on the CH$_{4}$/CO ratio}. \aap 562, A51.

\bibitem[{{Venot} et~al.(2015){Venot}, {H{\'e}brard}, {Ag{\'u}ndez}, {Decin},
  and {Bounaceur}}]{Venot2015}
{Venot}, O., {H{\'e}brard}, E., {Ag{\'u}ndez}, M., {Decin}, L., {Bounaceur},
  R., 2015. {New chemical scheme for studying carbon-rich exoplanet
  atmospheres}. \aap 577, A33.

\bibitem[{{Visscher} and {Fegley}(2005)}]{Visscher2005}
{Visscher}, C., {Fegley}, Jr., B., 2005. {Chemical Constraints on the Water and
  Total Oxygen Abundances in the Deep Atmosphere of Saturn}. \apj 623,
  1221--1227.

\bibitem[{{Visscher} et~al.(2010){Visscher}, {Moses}, and
  {Saslow}}]{Visscher2010}
{Visscher}, C., {Moses}, J.~I., {Saslow}, S.~A., 2010. {The deep water
  abundance on Jupiter: New constraints from thermochemical kinetics and
  diffusion modeling}. \icarus 209, 602--615.

\bibitem[{{Visscher} and {Moses}(2011)}]{Visscher2011}
{Visscher}, C., {Moses}, J.~I., 2011. {Quenching of Carbon Monoxide and Methane
  in the Atmospheres of Cool Brown Dwarfs and Hot Jupiters}. \apj 738, 72.

\bibitem[{{von Zahn} et~al.(1998){von Zahn}, {Hunten}, and
  {Lehmacher}}]{vonZahn1998}
{von Zahn}, U., {Hunten}, D.~M., {Lehmacher}, G., 1998. {Helium in Jupiter's
  atmosphere: Results from the Galileo probe helium interferometer experiment}.
  \jgr 103, 22815--22830.

\bibitem[{{Wakelam} et~al.(2012){Wakelam}, {Herbst}, {Loison}, {Smith},
  {Chandrasekaran}, {Pavone}, {Adams}, {Bacchus-Montabonel}, {Bergeat},
  {B{\'e}roff}, {Bierbaum}, {Chabot}, {Dalgarno}, {van Dishoeck}, {Faure},
  {Geppert}, {Gerlich}, {Galli}, {H{\'e}brard}, {Hersant}, {Hickson},
  {Honvault}, {Klippenstein}, {Le Picard}, {Nyman}, {Pernot}, {Schlemmer},
  {Selsis}, {Sims}, {Talbi}, {Tennyson}, {Troe}, {Wester}, and
  {Wiesenfeld}}]{Wakelam2012}
{Wakelam}, V., {Herbst}, E., {Loison}, J.-C., {Smith}, I.~W.~M.,
  {Chandrasekaran}, V., {Pavone}, B., {Adams}, N.~G., {Bacchus-Montabonel},
  M.-C., {Bergeat}, A., {B{\'e}roff}, K., {Bierbaum}, V.~M., {Chabot}, M.,
  {Dalgarno}, A., {van Dishoeck}, E.~F., {Faure}, A., {Geppert}, W.~D.,
  {Gerlich}, D., {Galli}, D., {H{\'e}brard}, E., {Hersant}, F., {Hickson},
  K.~M., {Honvault}, P., {Klippenstein}, S.~J., {Le Picard}, S., {Nyman}, G.,
  {Pernot}, P., {Schlemmer}, S., {Selsis}, F., {Sims}, I.~R., {Talbi}, D.,
  {Tennyson}, J., {Troe}, J., {Wester}, R., {Wiesenfeld}, L., 2012. {A KInetic
  Database for Astrochemistry (KIDA)}. \apjs 199, 21.

\bibitem[{{Wang} et~al.(2015){Wang}, {Gierasch}, {Lunine}, and
  {Mousis}}]{Wang2015}
{Wang}, D., {Gierasch}, P.~J., {Lunine}, J.~I., {Mousis}, O., 2015. {New
  insights on Jupiter's deep water abundance from disequilibrium species}.
  \icarus 250, 154--164.

\bibitem[{{Wang} et~al.(2016){Wang}, {Lunine}, and {Mousis}}]{Wang2016}
{Wang}, D., {Lunine}, J.~I., {Mousis}, O., 2016. {Modeling the
  disequilibrium species for Jupiter and Saturn: Implications for Juno and
  Saturn entry probe}. \icarus 276, 21--38.

\bibitem[{{Weiland} et~al.(2011){Weiland}, {Odegard}, {Hill}, {Wollack},
  {Hinshaw}, {Greason}, {Jarosik}, {Page}, {Bennett}, {Dunkley}, {Gold},
  {Halpern}, {Kogut}, {Komatsu}, {Larson}, {Limon}, {Meyer}, {Nolta}, {Smith},
  {Spergel}, {Tucker}, and {Wright}}]{Weiland2011}
{Weiland}, J.~L., {Odegard}, N., {Hill}, R.~S., {Wollack}, E., {Hinshaw}, G.,
  {Greason}, M.~R., {Jarosik}, N., {Page}, L., {Bennett}, C.~L., {Dunkley}, J.,
  {Gold}, B., {Halpern}, M., {Kogut}, A., {Komatsu}, E., {Larson}, D., {Limon},
  M., {Meyer}, S.~S., {Nolta}, M.~R., {Smith}, K.~M., {Spergel}, D.~N.,
  {Tucker}, G.~S., {Wright}, E.~L., 2011. {Seven-year Wilkinson Microwave
  Anisotropy Probe (WMAP) Observations: Planets and Celestial Calibration
  Sources}. \apjs 192, 19.

\bibitem[{{Wong} et~al.(2004){Wong}, {Mahaffy}, {Atreya}, {Niemann}, and
  {Owen}}]{Wong2004}
{Wong}, M.~H., {Mahaffy}, P.~R., {Atreya}, S.~K., {Niemann}, H.~B., {Owen},
  T.~C., 2004. {Updated Galileo probe mass spectrometer measurements of carbon,
  oxygen, nitrogen, and sulfur on Jupiter}. \icarus 171, 153--170.

\bibitem[{{Yung} et~al.(1988){Yung}, {Drew}, {Pinto}, and {Friedl}}]{Yung1988}
{Yung}, Y.~L., {Drew}, W.~A., {Pinto}, J.~P., {Friedl}, R.~R., 1988.
  {Estimation of the reaction rate for the formation of CH3O from H + H2CO -
  Implications for chemistry in the solar system}. \icarus 73, 516--526.

\end{thebibliography}


\end{document}